\documentclass[12pt]{article}

\usepackage{amssymb,amsthm}
\usepackage{graphicx}
\usepackage{fancyref}
\usepackage{multirow}
\usepackage{array}
\usepackage{amsfonts,amsmath}
\usepackage{longtable}

\usepackage{bm}
\usepackage{lmodern}
\usepackage{epsfig}
\usepackage{indentfirst}

\usepackage{overpic}
\usepackage{supertabular}

\usepackage{natbib}
\makeatletter                            
\renewcommand\@biblabel[1]{#1.}          
\makeatother %

\usepackage{url}
\usepackage{float}    
\usepackage{setspace}
\usepackage{enumerate}  
\usepackage{subfigure}   

\usepackage{extarrows}    

\usepackage[latin1]{inputenc}
\usepackage{tikz}
\usetikzlibrary{shapes,arrows,chains}
\usetikzlibrary{decorations.pathmorphing, backgrounds, positioning, fit, petri, automata}
\usepackage{xcolor}     

\definecolor{yellow1}{rgb}{1,0.8,0.2}      
\definecolor{LightBlue1}{RGB}{202,225,255}         
\definecolor{SteelBlue3}{RGB}{79,148,205}

\newcommand{\tabincell}[2]{\begin{tabular}{@{}#1@{}}#2\end{tabular}}

\newenvironment{sequation*}{\begin{equation*}\footnotesize}{\end{equation*}}

\newtheorem{theorem}{Theorem}
\newtheorem{lemma}{Lemma}

\begin{document}
\title{Testing normality using the summary statistics with application to meta-analysis}
\author{Dehui Luo$^{1}, $ Xiang Wan$^{2}$, Jiming Liu$^{2}$ and Tiejun Tong$^{1,}$\thanks{Corresponding author. E-mail: tongt@hkbu.edu.hk} \\ \\
{\small $^1$Department of Mathematics, Hong Kong Baptist University, Hong Kong}\\
{\small $^2$Department of Computer Science, Hong Kong Baptist University, Hong Kong}
}
\date{\today}
\maketitle

\begin{abstract}
\baselineskip 16pt
\noindent
As the most important tool to provide high-level evidence-based medicine, researchers can statistically summarize and combine data from multiple studies by conducting meta-analysis.  In meta-analysis, mean differences are frequently used effect size measurements to deal with continuous data, such as the Cohen's $d$ statistic \citep{cohen2013} and Hedges' $g$ statistic \citep{hedges1981} values.  To calculate the mean difference based effect sizes, the sample mean and standard deviation are two essential summary measures.  However, many of the clinical reports tend not to directly record the sample mean and standard deviation.  Instead, the sample size, median, minimum and maximum values and/or the first and third quartiles are reported.  As a result, researchers have to transform the reported information to the sample mean and standard deviation for further compute the effect size.  Since most of the popular transformation methods were developed upon the normality assumption of the underlying data, it is necessary to perform a pre-test before transforming the summary statistics.  In this article, we had introduced test statistics for three popular scenarios in meta-analysis.  We suggests medical researchers to perform a normality test of the selected studies before using them to conduct further analysis.  Moreover, we applied three different case studies to demonstrate the usage of the newly proposed test statistics.  The real data case studies indicate that the new test statistics are easy to apply in practice and by following the recommended path to conduct the meta-analysis, researchers can obtain more reliable conclusions.

\vskip 12pt
\noindent
\text{Keywords}: {Meta-analysis, Median, Mid-range, Mid-quartile range, Test statistic, Sample size}
\end{abstract}

\baselineskip 20pt
\newpage
\section{Introduction}
\noindent
Meta-analysis is the most important tool to provide high-level evidence in evidence-based medicine.  By conducting meta-analysis, researchers can statistically summarize and combine data from multiple studies with some pre-determined summary measure.  In most of the studies, mean difference is one of the frequently used effect size measurements for continuous data.  For example, Cohen's $d$ statistic \citep{cohen2013} and Hedges' $g$ statistic \citep{hedges1981} are two of the most famous mean difference measurements.  In order to calculate the mean difference based effect sizes, the sample mean and standard deviation are two essential summary measures.  In practical research, most of the medical studies provide the sample mean and standard deviation directly but, some of the clinical studies tend to report the summary statistics such as the sample median, quartiles and extremas.  As a result, researchers had developed a few methods to transform the reported information to the sample mean and standard deviation for further analysis.  In particular, \citet{hozo2005} was the first to establish estimators for the sample mean and standard deviation.  \citet{Tong2014} further improved Hozo et al.'s estimators of the sample standard deviation and \citet{luo2016} developed the optimal estimators of the sample mean.

These estimation methods, especially the methods proposed by Wan et al. were widely adopted in medical research area and had been frequently cited after published.  In particular, Wan et al.'s methods have already gained 274 citations in Google Scholar.  However, among the current meta-analysis studies, researchers usually apply the estimation methods to all kind of reported data directly, without considering the symmetry of original data.
For instance, we consider an example with the data obtained from \citet{Hawkins2017}, a meta-analysis about the association between B-type natriuretic peptides (BNP) and chronic obstructive pulmonary disease (COPD).  Totally there were 51 studies included in the meta-analysis and the authors discussed situations about stable disease cases, exacerbation cases and comparison between these two phases of patients.  For illustration purpose, we only reported 7 studies that recorded information of both patients and healthy individuals within stable disease phase in Table \ref{table_9}.  In \citet{Hawkins2017}, the authors used the methods in \citet{Tong2014} to estimate the sample mean and standard deviation.  The authors claimed that they found most of the recorded data were possibly not symmetric and they realized all the existing transformation methods were developed based on the symmetric assumption.  For example for Study 5, the median BNP level for the control group is 50 but the third quartile is just 51, which indicates that the underlying data are more likely to be right skewed.
In spite of this, the authors still conducted the transformation and they believed the major error of the statistical results were generated by the normality assumption of transformation methods.  Although Luo et al.'s methods of estimating the sample mean and Wan et al.'s standard deviation estimators were proved to have very good performance for both symmetric and skewed data, the authors in \citet{Hawkins2017} still had concern about the reliability of the transformation methods.  Under such circumstances, we believe it is essential to seek for a better solution to help people further filter the studies before performing the meta-analysis.

\renewcommand\arraystretch{1.5}            
\begin{table}[htp]\small
\tabcolsep 5pt \caption{\small Summary of included studies} \label{table_9} 
\begin{center}
\begin{tabular}{c|c|l|cc}  
  \hline
  \tabincell{c}{Index} &  \tabincell{c}{Study}  & & \tabincell{c}{Sample size}  & \tabincell{c}{BNP Levels}  \\   \hline

  \multirow{2}{*}{1} & \multirow{2}{*}{\citet{anderson2013}}       & \text{Case}    & 93   & 29$\pm$6      \\
                     &                                             & \text{Control} & 93   & 26(20-32)      \\
   \hline
  \multirow{2}{*}{2} & \multirow{2}{*}{\citet{gemici2008}}         & \text{Case}    & 17   & 21$\pm$16  \\
                     &                                             & \text{Control} & 17   & 13$\pm$11  \\
   \hline
  \multirow{2}{*}{3} & \multirow{2}{*}{\citet{boschetto2013}}      & \text{Case}    & 23   & 121(59-227)    \\
                     &                                             & \text{Control} & 23   & 50(43-51)     \\
   \hline
  \multirow{2}{*}{4} & \multirow{2}{*}{\citet{wang2013}}           & \text{Case}    & 80   & 245(196-336)  \\
                     &                                             & \text{Control} & 80   & 101(56-150)  \\
   \hline
  \multirow{2}{*}{5} & \multirow{2}{*}{\citet{beghe2013}}          & \text{Case}    & 70   & 115(50-364)      \\
                     &                                             & \text{Control} & 70   & 50(43-51)       \\
   \hline
  \multirow{2}{*}{6} & \multirow{2}{*}{\citet{bando1999}}          & \text{Case}    & 14   & 13$\pm$3      \\
                     &                                             & \text{Control} & 14   & 7$\pm$1       \\
  \hline
  \multirow{2}{*}{7} & \multirow{2}{*}{\citet{bozkanat2005}}       & \text{Case}    & 38   & 21$\pm$10      \\
                     &                                             & \text{Control} & 38   & 9$\pm$3       \\
  \hline
  \multicolumn{5}{l}{$\cdot$ Observations are expressed as \emph{mean $\pm$ SD} or \emph{median (interquartile range)}.}
\end{tabular}
\end{center}
\end{table}

\vskip 12pt
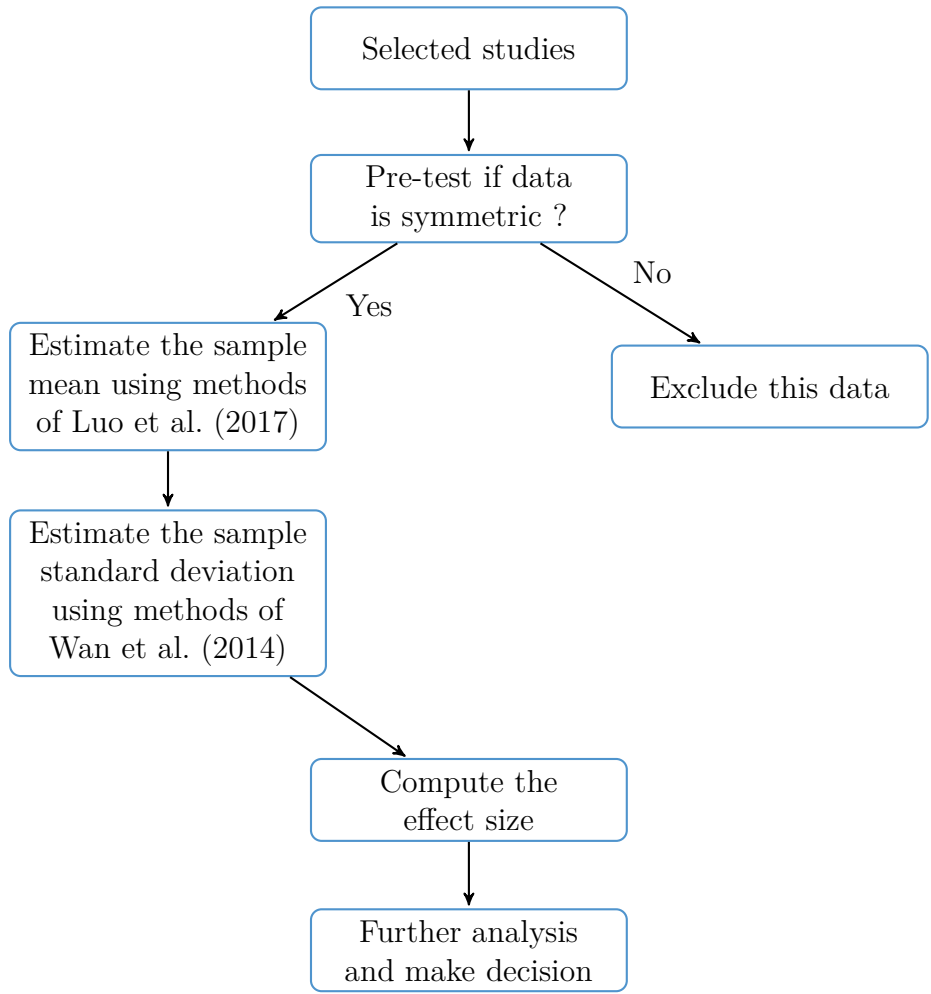
\begin{figure}[htp]
\centering
\begin{tikzpicture}[->,>=stealth',shorten >=1pt,auto,node distance=2.4cm,
                    thick]
  \tikzstyle{every state}=[rectangle,rounded corners,minimum height=6ex,fill=none,draw=SteelBlue3,text=black,text width=9.5em,text centered]

  \node[state]         (Start) at (0, 2)            {Selected studies};
  \node[state]         (test) at (0, 0)             {Pre-test if data is symmetric ?};
  \node[state]         (out1) at (-4, -2.5)         {Estimate the sample mean using methods of \citet{luo2016}};
  \node[state]         (out2) at (-4, -5.25)        {Estimate the sample standard deviation using methods of \citet{Tong2014}};
  \node[state]         (out3) at (4, -2.5)          {Exclude this data};
  \node[state]         (out4)  at (0, -8)         {Compute the effect size};
  \node[state]         (Decision) at (0, -10)      {Further analysis and make decision};

  \path (Start) edge    node {} (test)             
        (test)  edge    node {Yes} (out1)
        (test)  edge    node {No} (out3)
        (out1)  edge    node {} (out2)
        (out2)  edge    node {} (out4)
        (out4)   edge    node {} (Decision);
\label{fig:meta_procedure}
\end{tikzpicture}
\caption{Recommended procedure of computing effect sizes for meta-analysis}
\end{figure}

As a matter of fact, in both \citet{Tong2014} and \citet{luo2016}, the estimators of the sample mean and standard deviation are proposed based on the assumption that the underlying data is normally distributed.  Therefore, if the original data is skewed or very skewed, it might be inappropriate to treat this kind of data as normally distributed.  Furthermore, it is natural to consider that in clinical trial studies, when the underlying data is not symmetric, reporting the sample median rather than the sample mean is reasonable.  In this case, when the original data is skewed, transforming the reported information (such as the sample median, minimum and maximum values) to the sample mean and standard deviation may lead to lack of accuracy in the follow-up analysis.
In view of the above situation, we propose some new test statistics to pre-test whether the underlying data is normally distributed or not.  If there is no significant evidence to prove that the selected study is skewed, researchers may consider to transform the reported information to the sample mean and standard deviation.  Otherwise, we suggest that researchers may consider to not to include the tested study when conducting meta-analysis.  In this case, we suggest when conducting meta-analysis, researchers can follow the procedures as shown in the below flowchart.

Based on the above motivation, in this paper, we propose new test statistics for three most frequently used scenarios in clinical trial reports as mentioned in both \citet{Tong2014} and \citet{luo2016}, which may help researchers better choose the included studies for conducting meta-analysis.  For the proposed test statistics for each scenario, we conduct simulation studies to check whether the new test methods perform well in practice.  We also apply a few real data examples to illustrate the usefulness of the new test statistics.  Eventually, we summarize our new test statistics and discuss some future directions.

\
\section{Motivation}
To better illustrate the issue we mentioned in previous section, we choose five popular skewed distributions as examples to show how skewed data may affect the medical decision.  Suppose we are computing the effect size for some paired experiments about a certain disease.  For the sake of consistency, we follow the notations as used in \citet{luo2016}: using letters $a$, $m$, $b$ to denote the sample minimum value, median and maximum value for a study with size $n$, respectively.

In Table \ref{table_1}, let
the log-normal distribution with location parameter $\mu=0$ and scale parameter $\sigma=1$ for the disease cases, $\mu=1,\sigma=1$ for the controls;
the chi-square distribution with degrees of freedom $df=3$ for the disease cases, $df=4$ for controls;
the exponential distribution with rate parameter $\lambda=1$ for the disease cases, $\lambda=1.5$ for controls;
the beta distribution with shape parameters $\alpha=2$ and $\beta=5$ for the disease cases, $\alpha=2,\beta=7$ for controls;
and the Weibull distribution with shape parameter $k=1.5$ and scale parameter $\lambda=1$ for the disease cases, $k=3,\lambda=1$ for controls.
These distributions are used to generate the true sample mean, minimum ($a$), maximum ($b$) and median ($m$).  We then use the method in \citet{luo2016} to estimate the sample mean and use the method in \citet{Tong2014} to estimate the standard deviation for further analysis.  Eventually, we compute the effect size (Cohen's $d$ value) using the estimated sample mean as well as the actual sample mean and make comparison.  Note that for ease of computation, we assume the sample sizes for both the disease cases and controls are the same and the sample sizes for different distributions are arbitrarily chosen.

\renewcommand\arraystretch{1.5}
\begin{table*}[ht]\footnotesize
\tabcolsep 4pt \caption{\small Example of 5 skewed distributed data} \label{table_1}
\begin{center}
\begin{tabular}{lcccccc} 
  \hline
    \tabincell{c}{Data\\ Distribution}  &  \tabincell{c}{Size\\ (Cases)}  &  \tabincell{c}{Size (Controls)} & \tabincell{c}{Summary statistics \\(Cases)} &\tabincell{c}{Summary statistics \\(Controls)} \\   \hline
    Log-normal  & 350 & 350 & $a=0.03,m=1.00,b=28.11$   & $a=0.23,m=2.52,b=33.56$     \\
    Chi-square  & 200 & 200 & $a=0.06,m=2.66,b=9.60$    & $a=0.07,m=3.72,b=28.91$      \\
    Exponential & 150 & 150 & $a=0.002,m=0.67,b=6.82$   & $a=0.003,m=0.46,b=4.07$       \\
    Beta        & 300 & 300 & $a=0.001,m=0.26,b=0.76$   & $a=0.005,m=0.21,b=0.64$        \\
    Weibull     & 400 & 400 & $a=0.01,m=0.85,b=2.99$    & $a=0.16,m=0.86,b=1.93$         \\
  \hline
\end{tabular}
\end{center}
\end{table*}

\renewcommand\arraystretch{1.5}
\begin{table*}[ht]\footnotesize
\tabcolsep 5pt \caption{\small Effect sizes comparison of 5 skewed distributed data} \label{table_2}
\begin{center}
\begin{tabular}{lcccccc} 
  \hline
    \tabincell{c}{Data\\ Distribution}  & \tabincell{c}{(Cases) True\\ mean (SD)} & \tabincell{c}{(Controls) True\\ mean (SD)} & \tabincell{c}{(Cases)\\ Estimated \\ Mean (SD)}  & \tabincell{c}{(Controls)\\ Estimated \\ Mean (SD)} & \tabincell{c}{Effect Size \\(use estimated\\ mean)} &\tabincell{c}{Effect Size \\(use true\\ mean)} \\   \hline
    Log-normal  & 1.72 (2.70) & 4.05 (4.64)  & 1.61 (4.82) & 3.20 (5.72)  & 0.30   & 0.61  \\
    Chi-square  & 3.08 (2.17) & 4.71 (3.82)  & 2.81 (1.74) & 4.48 (5.27)  & 0.42   & 0.53 \\
    Exponential & 1.07 (1.13) & 0.70 (0.73)  & 0.90 (1.29) & 0.59 (0.12)  & -0.34  & -0.39 \\
    Beta        & 0.28 (0.16) & 0.23 (0.12)  & 0.27 (0.13) & 0.22 (0.11)  & -0.42  & -0.35 \\
    Weibull     & 0.93 (0.58) & 0.89 (0.33)  & 0.88 (0.50) & 0.87 (0.30)  & -0.02  & -0.08 \\
  \hline
\end{tabular}
\end{center}
\end{table*}

From Tables \ref{table_1} and \ref{table_2}, we clearly find that if the underlying distribution is very skewed, estimating the sample mean and standard deviation using the summary statistics may lead to an incorrect conclusion.  In particular, for the log-normal distribution and chi-square distribution in Table \ref{table_2}, the effect size computed by the actual sample mean is within the median effect level but the effect size computed by the estimated sample mean is in the low effect level.  As a result, testing the normality of the underlying data before estimating the sample mean and standard deviation is of crucial importance.  The normality test may help researchers to filter out the skewed data and further reduce the estimation error when computing the effect sizes.
Motivated by this circumstance, we will propose some new test statistics for three frequently used scenarios in Section 3. The simulation results and the real data case studies in the later sections indicate the good performance of the new test methods.

\

\section{Test methods}
\noindent
For the sake of consistency, we follow the same notations as those in \citet{luo2016}.  Let $n$ be the sample size and denote the 5-number summary for the data as
\begin{align}
&  a=\text{the minimum value},\nonumber \\
&  q_{1}=\text{the first quartile},\nonumber \\
&  m=\text{the median}, \nonumber\\
&  q_{3}=\text{the third quartile}, \nonumber\\
&  b=\text{the maximum value}. \nonumber
\end{align}

In this work, we consider the three most frequently occurred scenarios in clinic trial reports:
\begin{align*}
&  \mathcal{S}_{1}=\{a,m,b;n\},\nonumber \\
&  \mathcal{S}_{2}=\{q_{1},m,q_{3};n\}, \nonumber \\
&  \mathcal{S}_{3}=\{a,q_{1},m,q_{3},b;n\}. \nonumber
\end{align*}
According to \citep{triola2009bk}, we refer to $(b-a)$ as the range, $(a+b)/2$ as the mid-range, $(q_3-q_1)$ as the interquartile range,
and $(q_1+q_3)/2$ as the mid-quartile range.  Note that according to \citet{Tong2014} and \citet{luo2016}, the mid-range and the mid-quartile range are used to estimate the sample mean, while the range and the interquartile range are used to estimate the standard deviation.

For the following sections, our general assumption is letting $X_{1}, X_{2}, \ldots, X_{n}$ be a random sample of size $n$ from the normal distribution $N(\mu,\sigma^{2})$,
and $X_{(1)}\leq X_{(2)}\leq \cdots \leq X_{(n)}$ be the ordered statistics of the sample.
According to \citet{Chenhung}, we use $X_{([np])}$ to denote the sample $p$th quantile, where $p\in (0,1)$ and $[np]$ represents the integer part of $np$.  With the above notations, we have $a=X_{(1)}$, $q_1=X_{([0.25n])}$, $m=X_{([0.5n])}$, $q_3=X_{([0.75n])}$, and $b=X_{(n)}$.

For convenience, let also $X_{i}=\mu+\sigma Z_{i}$, or equivalently, $X_{(i)}=\mu+\sigma Z_{(i)}$ for $i=1, \ldots, n$.
Then $Z_{1}, Z_{2}, \ldots,Z_{n}$ follows the standard normal distribution $N(0,1)$,
and $Z_{(1)}\leq Z_{(2)}\leq \cdots \leq Z_{(n)}$ are the ordered statistics of the sample $\{Z_{1}, \ldots, Z_{n}\}$.

\

\subsection{Hypothesis test for $\mathcal{S}_{1}=\{a,m,b;n\}$}
\noindent
Note that the proposed estimators of the sample mean in \citet{luo2016} and the estimators of the standard deviation in \citet{Tong2014} are developed based on the normality assumption.  To ensure the accuracy of the estimation results, it is natural to test whether the population data are normally distributed or not.  That is, if the data does not pass the test, we may conclude the underlying data is not normally distributed and hence it is not appropriate to be used in the sample mean and standard deviation estimation.
Therefore, the hypothesis we proposed for this scenario is:
\begin{eqnarray*}
&&H_{0}: \text{The data follow a normal distribution},\\
&&H_{1}:\text{The data do not follow a normal distribution}.
\end{eqnarray*}
As the most important property of normal distribution is the symmetry property, we will use it to identify the normality of the underlying distribution.  Under the situation of scenario $\mathcal{S}_{1}$, only the sample median and extremes are reported.  Therefore, it is reasonable to consider comparing the distances between the sample median to the minimum as well as it to the maximum, i.e. computing the value of $(b-m)-(m-a)$.  If the difference is very close to zero, we may conclude that the original data is normally distributed.
Based on the above hypothesis, we introduce the following test statistic, $T$:
\begin{equation}
T=\frac{a+b-2m}{{\rm SE}(a+b-2m)}.
\label{eq:test_statistic_s1}
\end{equation}
By Theorem \ref{lemma:test_S1} in Appendix B, the simplified test statistic under the null hypothesis $H_{0}$ is
\begin{equation}
T_{0}=\frac{a+b-2m}{\sigma\sqrt{\frac{\pi^2}{6\log(n)}+\frac{\pi}{n}}}.
\label{eq:simple_test_s1}
\end{equation}
Note that $\sigma$ is unknown and need to be estimated.  Based on \citet{Tong2014}, the estimation of the sample standard deviation for scenario $\mathcal{S}_{1}$ is
\begin{equation}
\hat{\sigma}\approx \frac{b-a}{2\Phi^{-1}\left(\frac{n-0.375}{n+0.25}\right)}.
\label{eq:estimate_sigma_s1}
\end{equation}
Plugging (\ref{eq:estimate_sigma_s1}) into (\ref{eq:simple_test_s1}), we can easily get the modified test statistic for this scenario:
\begin{equation}
T_{1}=\frac{\tau(n)\left(a+b-2m\right)}{b-a},
\label{eq:final_test_s1}
\end{equation}
where $\tau(n)=2\Phi^{-1}\left(\frac{n-0.375}{n+0.25}\right)\Big/\sqrt{\frac{\pi^2}{6\log(n)}+\frac{\pi}{n}}$ is the coefficient function related to the sample size $n$.

Based on the above test statistic, if the observed $T_{1}$ is within the interval $[-1.96,1.96]$, we accept the null hypothesis and conclude that the underlying data is normally distributed.  Thus, we can continue to conduct the data transformation, which is to estimate the sample mean and standard deviation from the 5-number summary statistics.  Otherwise, if the null hypothesis is rejected, we suggest researchers not to include the tested data when proceeding meta-analysis.

\

\subsection{Hypothesis test for $\mathcal{S}_{2}=\{q_{1},m,q_{3};n\}$}
\noindent
Similar to the previous section, the hypothesis is defined as:
\begin{eqnarray*}
&&H_{0}: \text{The data follow a normal distribution},\\
&&H_{1}:\text{The data do not follow a normal distribution}.
\end{eqnarray*}
Under the case of scenario $\mathcal{S}_{2}$, only the sample quartiles are provided.
Based on the symmetry property of the normal distribution, in population aspect, the distance between the first quartile and the median is equal to that from the median to the third quartile.
As a result, we introduce a test statistic which compares these two distances, i.e. computing the value of $(q_{3}-m)-(m-q_{1})$.  By the above setting, the test statistic for this scenario is:
\begin{equation}
T=\frac{q_{1}+q_{3}-2m}{{\rm SE}(q_{1}+q_{3}-2m)}.
\label{eq:test_statistic_s2}
\end{equation}
By Theorem \ref{lemma:test_S2} in Appendix B, the simplified test statistic under the null hypothesis $H_{0}$ is
\begin{equation}
T_{0}= \frac{\sqrt{n}\left(q_{1}+q_{3}-2m\right)}{1.83\sigma}.
\label{eq:simple_test_s2}
\end{equation}
Similar to Section 2.1, $\sigma$ is unknown and by \citet{Tong2014}, the estimation of the sample standard deviation for scenario $\mathcal{S}_{2}$ is
\begin{equation}
\hat{\sigma}\approx\frac{q_{3}-q_{1}}{2\Phi^{-1}\left(\frac{0.75n-0.125}{n+0.25}\right)}.
\label{eq:estimate_sigma_s2}
\end{equation}
Plugging (\ref{eq:estimate_sigma_s2}) into (\ref{eq:simple_test_s2}), we can easily obtain the updated test statistic for this scenario:
\begin{equation}
T_{2}=\frac{\varphi(n)\left(q_{1}+q_{3}-2m\right)}{q_{3}-q_{1}},
\label{eq:final_test_s2}
\end{equation}
where $\varphi(n)=1.09\sqrt{n}\Phi^{-1}\left(\frac{0.75n-0.125}{n+0.25}\right)$ is the coefficient function related to the sample size $n$.

Similarly, if the observed $T_{2}$ is within the interval $[-1.96,1.96]$, we accept the null hypothesis and regard the underlying data as applicable to perform the estimation of the sample mean and standard deviation.  Otherwise, we suggest that researchers should remove the tested data from further analysis procedure.

\subsection{Test statistic for $\mathcal{S}_{3}=\{a,q_{1},m,q_{3},b;n\}$}
\noindent
In this section, we discuss the case of all 5-number summary statistics are provided.  Similar to the previous two subsections, the hypothesis is defined as:
\begin{eqnarray*}
&&H_{0}: \text{The data follow a normal distribution},\\
&&H_{1}:\text{The data do not follow a normal distribution}.
\end{eqnarray*}
Recall that in the previous subsections, the symmetry property of normal distribution is used to test the normality of underlying data.
In Section 2.1, the test statistic is built by comparing the difference in distances between the sample extremes to the median, i.e. computing $(b-m)$ and $(m-a)$, respectively. In Section 2.2, the value of $(q_{3}-m)-(m-q_{1})$ is computed to measure the difference in distances between the sample quartiles to the median.  Hence, taking both previous ideas into account, we introduce the following test statistic:
\begin{equation}
T=\frac{a+b+q_{1}+q_{3}-4m}{{\rm SE}(a+b+q_{1}+q_{3}-4m)}.
\label{eq:test_statistic_s3}
\end{equation}
By Theorem \ref{lemma:test_S3} in Appendix B, the simplified test statistic under the null hypothesis is
\begin{equation}
T_{0}= \frac{a+b+q_{1}+q_{3}-4m}{\sigma\sqrt{\frac{\pi^2}{6\log(n)} + \frac{10.14}{n}}},
\label{eq:simple_test_s3}
\end{equation}
where $\sigma$ is unknown.  Therefore, the next task is to determine an estimator of the sample standard deviation $\sigma$.

Following the similar idea as in \citet{Tong2014}, by Lemma \ref{theorem:order_statistics}, we have $E(b-a+q_{3}-q_{1})=2\sigma\left[E(Z_{n})+E_(Z_{([0.75n])})\right]$, which leads to the estimator $\hat{\sigma}\approx (b-a+q_{3}-q_{1})/\left(2E(Z_{(n)})+2E(Z_{([0.75n])})\right)$.  According to Blom's method of approximating $E(Z_{i})$ \citep{blom1958}, the estimator of the sample standard deviation is defined as
\begin{equation}
\hat{\sigma} \approx \frac{b-a+q_{3}-q_{1}}{2\Phi^{-1}\left(\frac{n-0.375}{n+0.25}\right)+2\Phi^{-1}\left(\frac{0.75n-0.125}{n+0.25}\right)}.
\label{eq:estimate_sigma_s3}
\end{equation}
Consequently, plugging (\ref{eq:estimate_sigma_s3}) into (\ref{eq:simple_test_s3}), the updated test statistic $T_{3}$ has the following expression:
\begin{equation}
T_{3}=\frac{\kappa(n)(a+b+q_{1}+q_{3}-4m)}{b-a+q_{3}-q_{1}},
\label{eq:final_test_s3}
\end{equation}
where $\kappa(n)=\left[2\Phi^{-1}\left(\frac{n-0.375}{n+0.25}\right)+2\Phi^{-1}\left(\frac{0.75n-0.125}{n+0.25}\right)\right]\Big/\sqrt{\frac{\pi^2}{6\log(n)} + \frac{10.5}{n}}$ is the coefficient function related to the sample size $n$.

Based on the above definition, if the observed $T_{3}$ is within the interval $[-1.96,1.96]$, we accept the null hypothesis.  Otherwise, the null hypothesis is rejected and we suggest researchers not to include the tested data into meta-analysis.

\

\section{Simulation studies}
\noindent
In this section, we conduct some simulation studies to evaluate the performance of the test statistics in Section 2.  The type I error and the statistical power will be computed.  To compute the statistical power of test statistics (\ref{eq:final_test_s1}) and (\ref{eq:final_test_s2}), we choose the following skewed distributions as the alternative distribution: the standard log-normal distribution with $\mu=0,\sigma=1$; the standard exponential distribution with $\lambda=1$; the beta distribution with shape parameters $\alpha=1,\beta=5$; the Chi-square distribution with degree of freedom $df=1$ and the Weibull distribution with scale parameter $\lambda=1$, shape parameter $k=2$.

The simulation results indicate that all three test statistics can provide an acceptable statistical power, as well as the type I error under control.

\

\subsection{Simulation study for $\mathcal{S}_{1}$}
\begin{figure}[htp]
 \begin{center}
 \includegraphics[scale=.7]{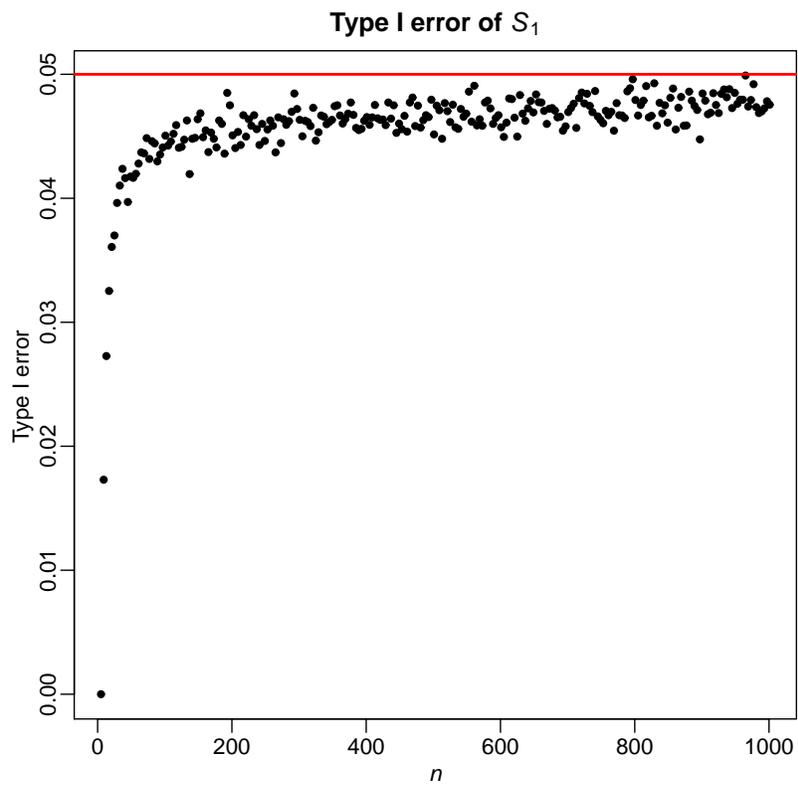}
 \end{center}
 \caption{Type I error for scenario $\mathcal{S}_1$.}
 \label{fig:type1_error_s1}
 \end{figure}

\begin{figure}[htp]
 \begin{center}
 \includegraphics[scale=.8]{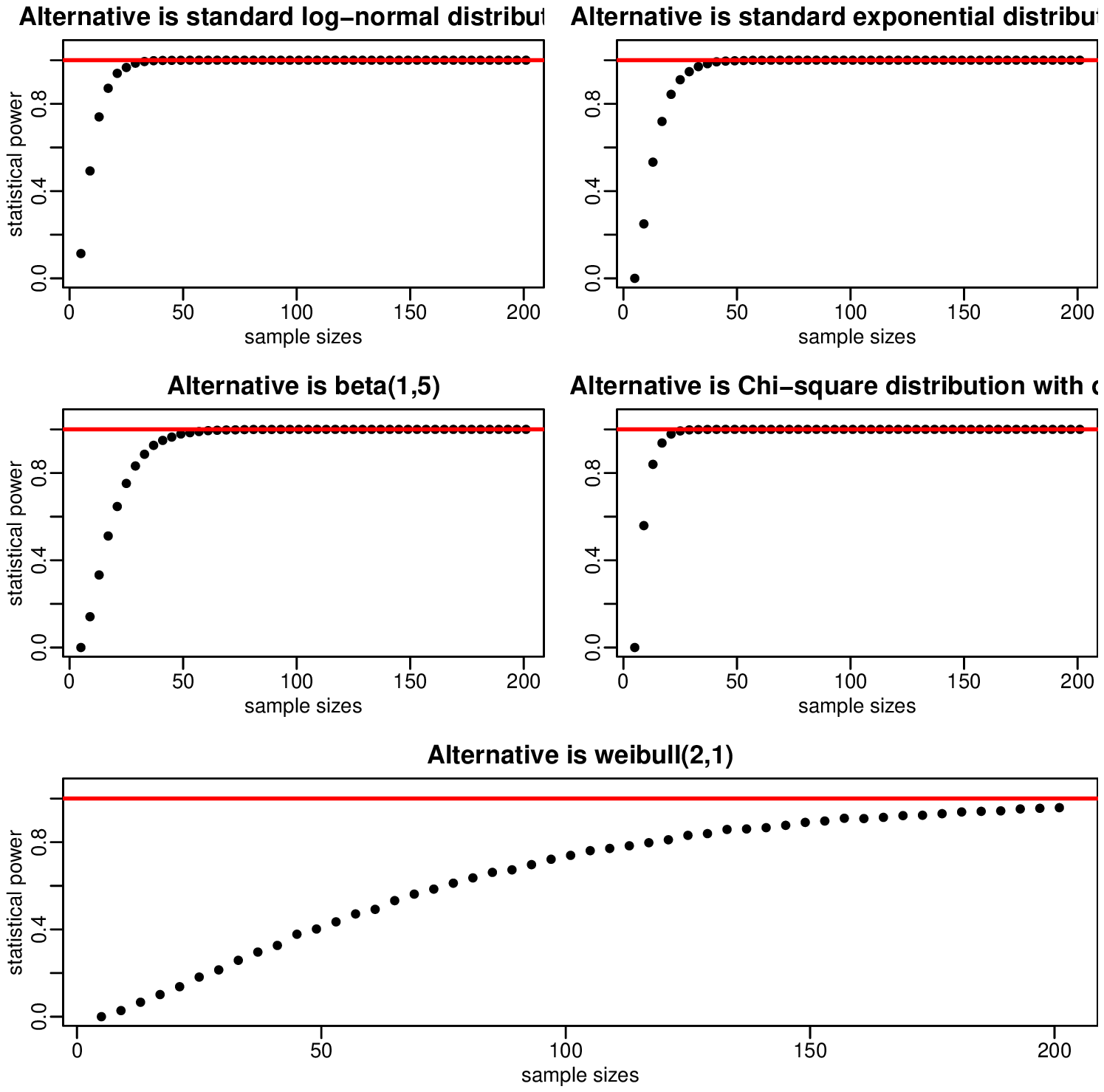}
 \end{center}
 \caption{The statistical power for scenario $\mathcal{S}_1$.}
 \label{fig:power_s1}
 \end{figure}

\noindent
Figure \ref{fig:type1_error_s1} shows the type I error of test statistic $T_{1}$ in scenario $\mathcal{S}_{1}$ (Eq. (\ref{eq:final_test_s1})).  Since in practical research, the sample size $n$ rarely drop below 30, we tend to focus on the situation when $n$ is large.  Hence, although the type I error of $T_{1}$ is not very close to 0.05 for small sample sizes, we consider it acceptable because the type I error is able to maintain a value around 0.05 as $n$ increases to 200 or larger.  Figure \ref{fig:power_s1} reports the statistical power of test statistic $T_{1}$. The simulation studies were conducted by assuming the alternative distribution is not normal distribution.  It is obvious that for all the five skewed distributions, the statistical power increase to 1 rapidly (mostly increase to 1 before $n$ reaches 100).  Consequently, with the type I error very close to 0.05 and a statistical power close to 1, we consider that test statistic $T_{1}$ may have impressive performance in practical application and we will conduct a real data analysis to evaluate its performance in the next section.

\

\subsection{Simulation study for $\mathcal{S}_{2}$}
\begin{figure}[htp]
 \begin{center}
 \includegraphics[scale=.7]{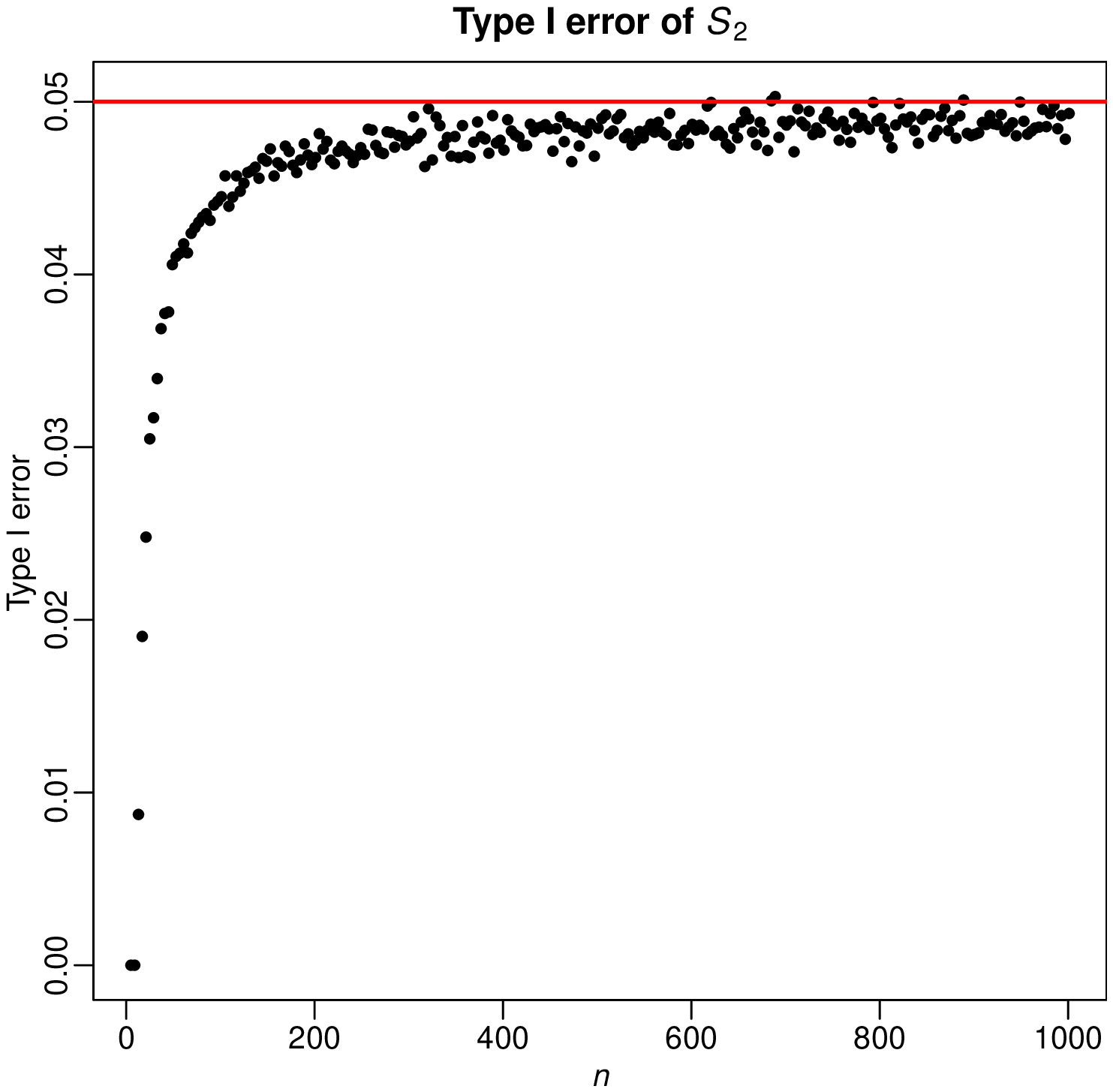}
 \end{center}
 \caption{Type I error for scenario $\mathcal{S}_2$.}
 \label{fig:type1_error_s2}
 \end{figure}

\begin{figure}[htp]
 \begin{center}
 \includegraphics[scale=.8]{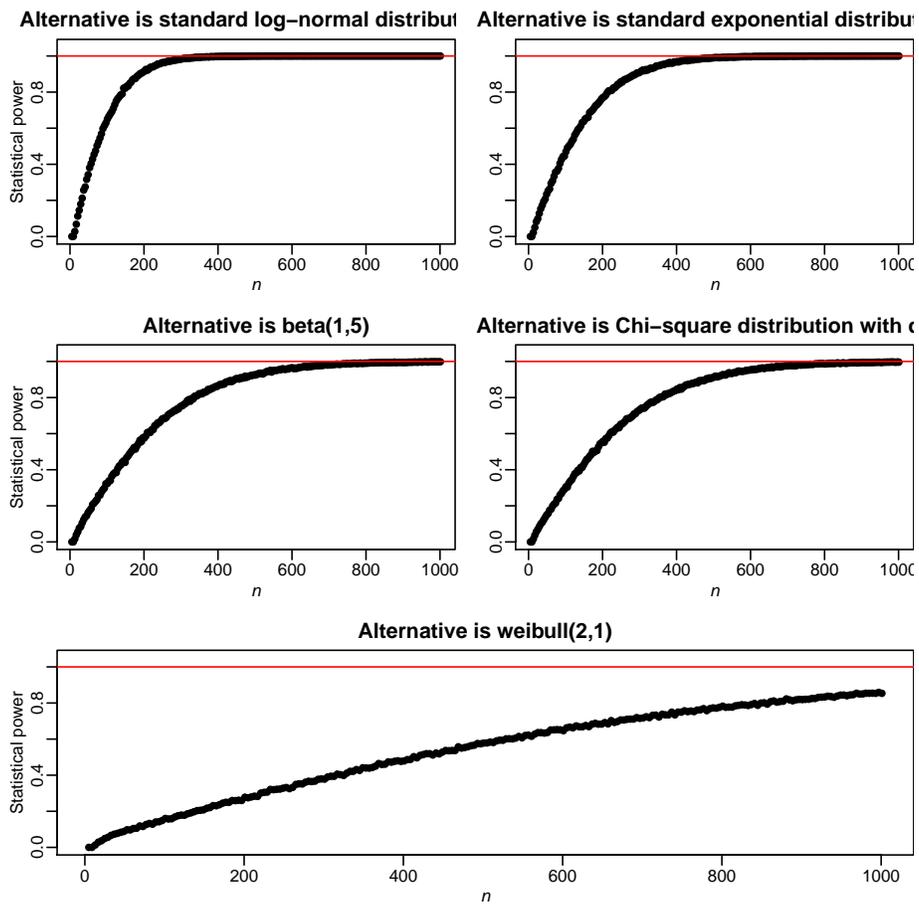}
 \end{center}
 \caption{The statistical power for scenario $\mathcal{S}_2$.}
 \label{fig:power_s2}
 \end{figure}

\noindent
Similar to section 3.1, Figure \ref{fig:type1_error_s2} shows the type I error of test statistic $T_{2}$ in scenario $\mathcal{S}_{2}$ (Eq. (\ref{eq:final_test_s2})).  It is evident that the type I error is able to maintain a value around 0.05 especially when the sample size $n$ increases to more than 200.  Figure \ref{fig:power_s2} reports the statistical power of test statistic $T_{2}$ when the alternative distributions are skewed distributions.  Compare to section 3.1, the statistical power of test statistic $T_{2}$ increases to 1 in a slower motion, it reaches 1 only when the sample size $n$ is larger than 400.  However, since in practice, most of the studies have large sample sizes, it is acceptable that test statistic $T_{2}$ may provide a statistical power that reaches 1 in a slightly slower speed.
Also, as all the value of type I error is around 0.05, we consider test statistic $T_{2}$ may have a very good performance in real-life application.  In the next section, we will conduct real data analysis to evaluate the performance of $T_{2}$.

\subsection{Simulation study for $\mathcal{S}_{3}$}
\begin{figure}[htp]
 \begin{center}
 \includegraphics[scale=.7]{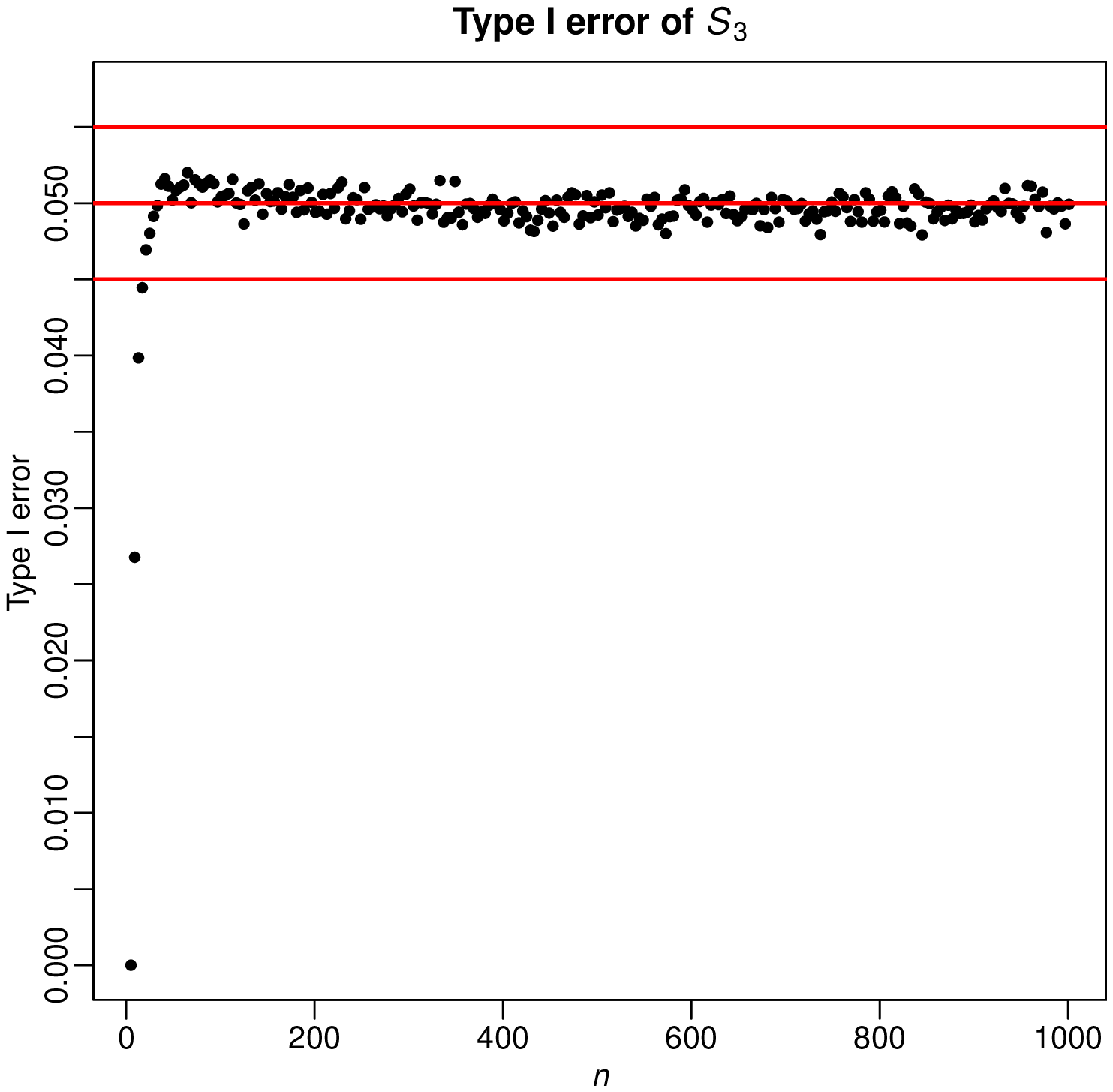}
 \end{center}
 \caption{Type I error for scenario $\mathcal{S}_3$.}
 \label{fig:type1_error_s3}
 \end{figure}

\begin{figure}[htp]
 \begin{center}
 \includegraphics[scale=.8]{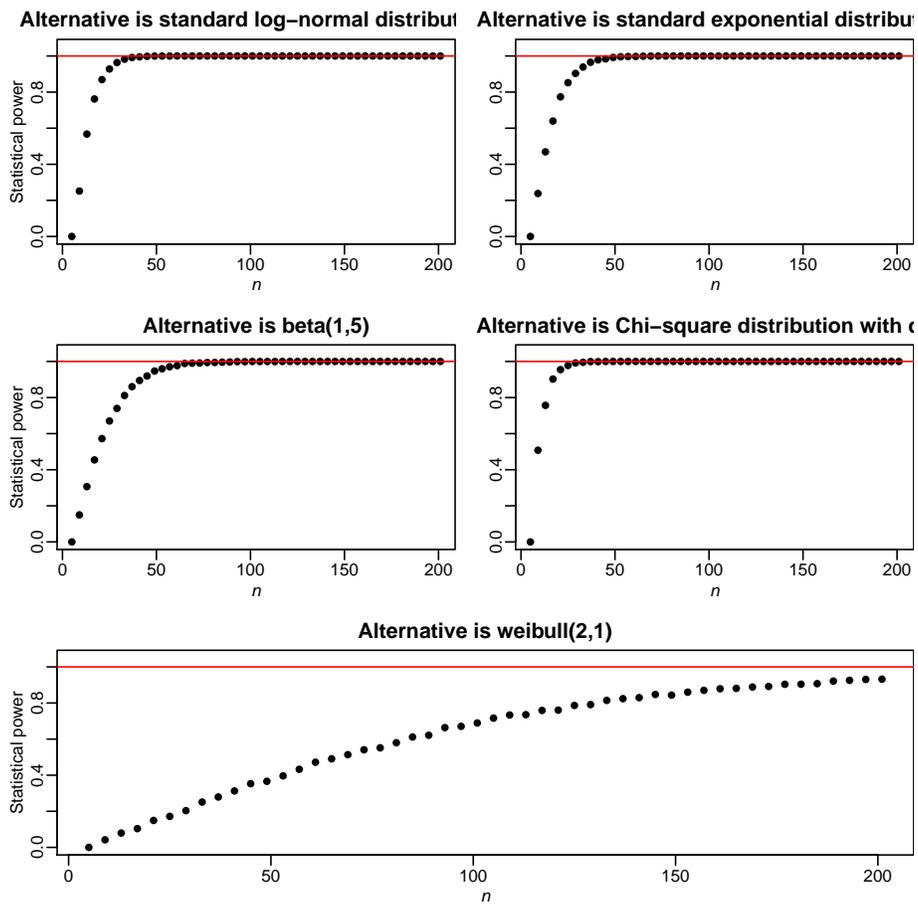}
 \end{center}
 \caption{The statistical power for scenario $\mathcal{S}_3$.}
 \label{fig:power_s3}
 \end{figure}

\noindent
In this section, Figure \ref{fig:type1_error_s3} shows the type I error of test statistic $T_{3}$ in scenario $\mathcal{S}_{3}$ (Eq. (\ref{eq:final_test_s3})).  In Figure \ref{fig:type1_error_s3}, it is obvious that the type I error of $T_{3}$ drop within the range of [-0.045,0.055] and most of the points are very close to 0.05.  Figure \ref{fig:power_s3} reports the statistical power of test statistic $T_{3}$ with skewed alternative distributions.  Similar to scenario $\mathcal{S}_{1}$, the statistical power of $T_{3}$ reaches toward 1 rapidly.  Therefore, with a type I error around 0.05 and statistical power close to 1, we expect that $T_{3}$ would have very good performance in practice.

\

\section{Real data analysis}
\noindent
In this section, we tend to apply 2 real data analysis as examples to demonstrate the usage of our proposed test statistics.  The first case is about investigating the association between asthma and leptin and adiponectin serum levels, respectively \citep{zhang2017}.  The second case is to identify the effects of statin therapy on four indicators of plasma lipid concentrations in HIV-infected patients \citep{banach2016}.  In both cases, some data are recorded as the sample median and interquartile ranges, or median with the sample extremum.  In this section, test statistics $T_{1}$ (\ref{eq:final_test_s1}) and $T_{2}$ (\ref{eq:final_test_s2}) are used to conduct the symmetry test of underlying data.

\

\subsection{Case study of investigating the association of asthma diagnosis with leptin and adiponectin}
\noindent
The first data obtained from a meta-analysis about the effects of leptin and adiponectin serum levels on the diagnosis of asthma \citep{zhang2017}.
The article is published on the Journal of Investigative Medicine (JIM), one of the BMJ journals.  It includes 13 studies and all the analysis is divided into 2 parts, one focus on leptin serum level and the other focus on adiponectin serum level.  For leptin serum level, 5 of the studies report the sample median and interquartile range (satisfies scenario $\mathcal{S}_{2}$), 1 reports median, minimum and maximum values (satisfies scenario $\mathcal{S}_{1}$).  For adiponectin serum level, two of the studies report the sample median and interquartile range.
Hence, in this section, test statistics $T_{1}$ and $T_{2}$ will be applied to evaluate whether the underlying data is symmetry and can be further used to estimate the sample mean and standard deviation.

\

\subsubsection{Data description}
\noindent
The reported information of the 13 studies in \citet{zhang2017} are displayed in the following Table \ref{table_3}.  There are 12 studies that record the values of leptin serum level and 11 studies that record the values of adiponectin serum level.  Since for both leptin and adiponectin serum levels, some of the selected studies only provide summary measurements (the sample median, interquartile range or extremum), it is essential to estimate the sample mean and standard deviation (SD) from recorded information.  However, all the existing sample mean and SD estimation methods are developed based on normality assumption.  As a result, before conducting the transformation, we will apply the symmetry test as proposed in Section 2 to check whether the selected studies are reasonable to be transformed.

\renewcommand\arraystretch{1.5}
\begin{table}[htp]\footnotesize
\tabcolsep 3pt \caption{Summary of included studies in \citet{zhang2017}} \label{table_3} \vspace*{-8pt}
\begin{center}
\begin{tabular}{@{\extracolsep{\fill}}c|l|ccc}  
  \hline
  \text{Study}  & \tabincell{c}{Type of asthma} & \tabincell{c}{Sample size}  &\text{Leptin}   & \text{Adiponection}  \\   \hline

   \multirow{2}{*}{\citet{hayashikawa2015}}     & \text{Asthma}             & 23  & \text{NS}           & 13.47$\pm$9.08   \\
                                                & \text{Healthy}            & 68  & \text{NS}           & 14.04$\pm$8.82   \\
   \hline
   \multirow{2}{*}{\citet{haidari2014}}         & \text{Asthma}             & 47  & 1.41$\pm$0.50       & 6.68$\pm$2.07    \\
                                                & \text{Healthy}            & 47  & 0.59$\pm$0.19       & 7.55$\pm$2.10    \\
   \hline
   \multirow{2}{*}{\citet{sood2014}}            & \text{Asthma}             & 44  & 34032.8$\pm$27597.8     & 4180.6$\pm$2671.1 \\
                                                & \text{Healthy}            & 44  & 33263.4$\pm$27874.9     & 3987.4$\pm$3106.0  \\
   \hline
   \multirow{2}{*}{\citet{cobanoglu2013}}       & \text{Asthma}             & 23  & 5.3[0.4-27.4]       & \text{NS}        \\
                                                & \text{Healthy}            & 51  & 8.8[0.3-31.3]       & \text{NS}      \\
   \hline
   \multirow{2}{*}{\citet{tsaroucha2013}}       & \text{Asthma}             & 32  & 24.8$\pm$14.8       & 13.5$\pm$9.2 \\
                                                & \text{Healthy}            & 22  & 13.7$\pm$10.0       & 10.1$\pm$6.4 \\
   \hline
   \multirow{3}{*}{\citet{yuksel2012}}          & \text{Obese asthma}       & 40  & 11.8$\pm$7.9        & 12586.2$\pm$3724.1    \\
                                                & \text{Non-obese asthma}   & 51  & 5.3$\pm$6.8         & 18089.3$\pm$6452.3 \\
                                                & \text{Healthy}            & 20  & 2.1$\pm$2.4         & 20297.5$\pm$3680.7    \\
   \hline
   \multirow{2}{*}{\citet{sideleva2012}}        & \text{Asthma}             & 11  & 0.3051$\pm$0.047     & 0.3471$\pm$0.037   \\  
                                                & \text{Healthy}            & 15  & 0.1256$\pm$0.016     & 0.8666$\pm$0.134   \\
   \hline
   \multirow{2}{*}{\citet{daSilva2012}}         & \text{Asthma}             & 26  & 38(30-60)           & 4.5(3.5-8.5)   \\     
                                                & \text{Healthy}            & 50  & 39(25-50)           & 4(3-7.8)   \\
   \hline
   \multirow{2}{*}{\citet{giouleka2011}}        & \text{Asthma}             & 100 & 9.6(7.6-16.25)      & 6.2(5.4-7.3)    \\
                                                & \text{Healthy}            & 60  & 7.2(4.6-10.3)       & 8.2(5.8-13.5)    \\
   \hline
   \multirow{2}{*}{\citet{leivo2011}}           & \text{Asthma}             & 35  & 0.5(0.5-1.1)        & 165$\pm$9.5 \\       
                                                & \text{Healthy}            & 32  & 0.6(0.4-0.8)        & 176$\pm$13 \\
   \hline
   \multirow{2}{*}{\citet{jang2009}}            & \text{Asthma}             & 60  & 2.31$\pm$0.04       & 1.90$\pm$0.17   \\
                                                & \text{Healthy}            & 30  & 2.22$\pm$0.06       & 1.95$\pm$0.04 \\
   \hline
   \multirow{3}{*}{\citet{kim2008}}             & \text{Atopic asthma}      & 149 & 2.27(0.65-5.03)     & 7.60$\pm$3.84       \\
                                                & \text{Non-atopic asthma}  & 37  & 2.22(0.96-3.29)     & 8.10$\pm$4.73       \\
                                                & \text{Healthy}            & 54  & 2.10(0.71-4.49)     & 7.32$\pm$4.19       \\
   \hline
   \multirow{2}{*}{\citet{guler2004}}           & \text{Asthma}             & 102 & 3.53(2.06-7.24)     & \text{NS}       \\
                                                & \text{Control}            & 33  & 2.26(1.26-4.71)     & \text{NS}       \\
  \hline
  \multicolumn{5}{l}{$\cdot$ Observations are expressed as mean $\pm$ SD, median (interquartile range) or median [minimum-maximum].}\\
  \multicolumn{5}{l}{$\cdot$ $NS$ indicates the information is not specified in the original study.}
\end{tabular}
\end{center}
\end{table}

\

\subsubsection{Results of symmetry test and meta-analysis}
\noindent
The results of the symmetry tests are recorded in Table \ref{table_4}.  Note that test statistic $T_{1}$ is applied on study \citet{cobanoglu2013} as it reports the sample median and extremum, test statistic $T_{2}$ is applied to studies \citet{daSilva2012}, \citet{giouleka2011}, \citet{leivo2011}, \citet{kim2008} and \citet{guler2004}.  To further proceed meta-analysis, it is more reliable to only transform the studies with symmetric data, to the sample mean and standard deviation via methods proposed by \citet{luo2016} and \citet{Tong2014}.  Thus, if the studies in Table \ref{table_4} have p-value greater than 0.05 on both asthma samples and healthy samples, these studies can be used to estimate the sample mean and standard deviation.  Note also that in \citet{kim2008}, patients were divided into three groups: obese asthma, non-obese asthma and healthy individuals.  In this case, only when the p-values of all these three groups are greater than 0.05, we would compute the sample mean and standard deviation for \citet{kim2008} and further combine the information of the first two groups.

\renewcommand\arraystretch{1.5}
\begin{table}[htp]\footnotesize
\tabcolsep 5pt \caption{\small Results of normality test on reported data from \citet{zhang2017}} \label{table_4} \vspace*{-8pt}
\begin{center}
\begin{tabular}{c|l|c|l|l|l|l}  
  \hline
  \multirow{3}{*}{Study}  & \multirow{3}{*}{\tabincell{c}{Type of asthma}} & \multirow{3}{*}{\tabincell{c}{Sample \\ size}}  &\multicolumn{2}{c|}{Leptin}   & \multicolumn{2}{c}{Adiponection}  \\   \cline{4-7}
                                    &  &  &\multirow{2}{*}{\tabincell{c}{Test\\ statistic}} & \multirow{2}{*}{p-value}   &\multirow{2}{*}{\tabincell{c}{Test\\ statistic}} & \multirow{2}{*}{p-value} \\
                                                 & & & & & &\\
   \hline
   \multirow{2}{*}{\citet{cobanoglu2013}}       & \text{Asthma}             & 23  & 3.022   & 0.003     & NS        & NS      \\
                                                & \text{Healthy}            & 51  & 2.935   & 0.003     & NS        & NS      \\
   \hline
   \multirow{2}{*}{\citet{daSilva2012}}         & \text{Asthma}             & 26  & 1.653   & 0.098     & 2.126     & 0.034   \\     
                                                & \text{Healthy}            & 50  & -0.606  & 0.545     & 2.945     & 0.003   \\
   \hline
   \multirow{2}{*}{\citet{giouleka2011}}        & \text{Asthma}             & 100 & 3.895   & $<0.001$  & 1.144     & 0.253   \\
                                                & \text{Healthy}            & 60  & 0.488   & 0.626     & 2.093     & 0.036   \\
   \hline
   \multirow{2}{*}{\citet{leivo2011}}           & \text{Asthma}             & 35  & 4.171   & $<0.001$  & NS        & NS \\   
                                                & \text{Healthy}            & 32  & 2.205$\times 10^{-15}$  & 1   & NS  & NS \\
   \hline
   \multirow{3}{*}{\citet{kim2008}}             & \text{Atopic asthma}      & 149 & 2.313   & 0.021     & NS       & NS \\
                                                & \text{Non-atopic asthma}  & 37  & -0.351  & 0.726     & NS       & NS \\
                                                & \text{Healthy}            & 54  & 1.391   & 0.164     & NS       & NS \\
   \hline
   \multirow{2}{*}{\citet{guler2004}}           & \text{Asthma}             & 102 & 3.189   & 0.001     & NS       & NS \\
                                                & \text{Control}            & 33  & 1.697   & 0.090     & NS       & NS \\
  \hline
  \multicolumn{5}{l}{$\cdot$ \emph{p-value}$<0.05$ indicates the underlying data is not symmetric.}  \\
  \multicolumn{5}{l}{$\cdot$ \emph{NS} indicates the reported data do not need to conduct the symmetry test.}
\end{tabular}
\end{center}
\end{table}

It is obvious that for leptin serum level, all the studies, except for study \citet{daSilva2012}, have p-values less than 0.05 on either asthma samples or healthy samples.  That is, these 5 studies (\citet{cobanoglu2013},\citet{giouleka2011}, \citet{leivo2011}, \citet{kim2008} and \citet{guler2004}) is not suitable to further conduct the sample mean and standard deviation estimation.  We suggest to exclude these five studies from the meta-analysis of investigating the association between leptin serum level and asthma.

 \begin{figure}[htp]
 \begin{center}
 \subfigure[Leptin level]{
 \label{fig:case_1_leptin_forest}
 \includegraphics[width=6.2in]{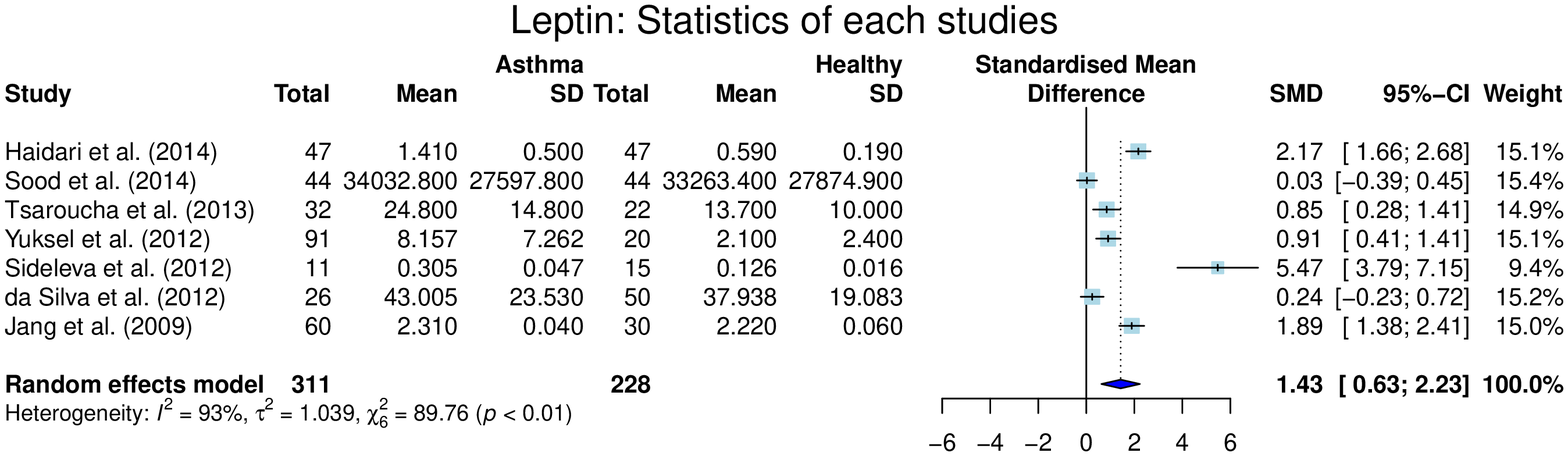}}
 \subfigure[Adiponectin level]{
 \label{fig:case_1_adiponectin_forest}
 \includegraphics[width=6.2in]{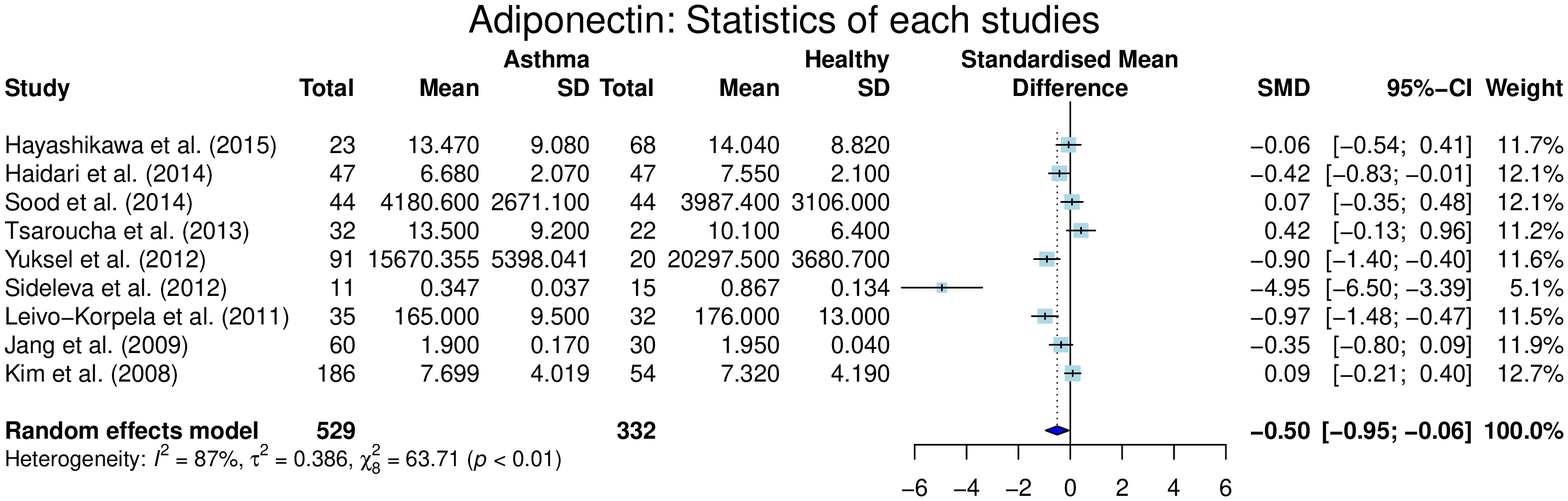}}
 \end{center}
 \caption{Forest plots for association between asthma versus leptin (\ref{fig:case_1_leptin_forest}) and adiponectin levels (\ref{fig:case_1_adiponectin_forest}), respectively.}
 \label{fig:case_1_forest_plots}
 \end{figure}

Note that in studies \citet{yuksel2012} and \citet{kim2008}, asthma patients were separated into two different levels.  Therefore, before computing the standardized mean difference (SMD, i.e. the Cohen's $d$ value \citep{cohen2013}) as the effect size for these two studies, we need to combine the information of those two levels of asthma as one.  Note also that since there are 5 studies did not pass the symmetry test, we excluded those studies in the meta-analysis.  As a result, we have the following forest plots for leptin and adiponectin levels.  During computation, we used \citet{luo2016}'s methods to estimate the sample mean value and \citet{Tong2014}'s methods to estimate the standard deviations for studies that only provided median and interquartile range or extremum.  Based on Figure \ref{fig:case_1_forest_plots}, there is extreme heterogeneity among the studies for both leptin and adiponectin levels ($Q=89.76, I^{2}=93\%, p<0.01$ for leptin; $Q=63.71, I^{2}=87\%, p<0.01$ for adiponectin).  The overall results indicate that patients with asthma have significant higher levels of leptin than healthy individuals (pooled standardized mean difference (SMD)$=1.43$, with $95\%$ CI from 0.63 to 2.23), and lower levels of adiponectin than healthy individuals (pooled SMD$=-0.50$, with $95\%$ CI from -0.95 to -0.06).  Since this section is aimed to provide a demonstration of how to use the proposed test statistics before conducting meta-analysis, we would not further proceed the sensitivity analysis and the publication bias analysis.

The pooled conclusion is similar to \citet{zhang2017}.  However, in the original article, the authors did not conduct the symmetry test and they used Hozo et al.'s methods\citep{hozo2005} to estimate the sample means and standard deviations.  As we had mention in introduction section, blindly estimate the sample mean and standard deviation before testing the symmetry of underlying data may yield an inaccurate conclusion.  Therefore, we suggest researchers to follow the procedure in Figure \ref{fig:meta_procedure}: conduct symmetry test for selected studies, exclude studies that may have skewed underlying data and then compute the individual and pooled effect sizes.

\

\subsection{Case study of investigating the effects of statin therapy on plasma lipid concentrations in HIV-infected patients}
\noindent
The second data is a meta-analysis to investigate the effects of statin therapy on plasma lipid concentrations in HIV-infected patients \citep{banach2016}. In this article, meta-analysis contains 12 randomized control trials (RCT) with 697 participants.  The reported data included 5 types of cholesterol level that may influence plasma concentration: low density lipoprotein cholesterol (LDL-C), total cholesterol, high density lipoprotein cholesterol (HDL-C), non-high density lipoprotein cholesterol (non-HDL-C) and triglycerides.  Note that in \citet{banach2016}, only two studies reports data of non-HDL-C and both these two studies provided the sample mean and standard deviation.  Hence in the following discussion, we will remove variable non-HDL-C and focus on the other four types of cholesterol level.

\

\subsubsection{Data description}
\noindent
The data of included studies are displayed in the following Table \ref{table_5}.  For each type of cholesterol level, data are reported either as the sample mean with standard deviation, or as median with interquartile range.  In this case, the test statistic $T_{2}$ will be used to conduct the symmetry test of the underlying data.  Similar to the previous case study, selected studies that have p-value greater than 0.05 are recommended to further proceed the sample mean and standard deviation estimation.  Otherwise, we regard that there is evidence to conclude the underlying data of selected studies are not symmetric and hence, we suggest researchers to remove those studies from the meta-analysis in order to obtain more precise conclusion.

\renewcommand\arraystretch{1.5}            
\begin{table}[htp]\footnotesize
\tabcolsep 4pt \caption{\small Summary of included studies in \citet{banach2016}} \label{table_5} 
\begin{center}
\begin{tabular}{@{\extracolsep{\fill}}c|l|ccccc}  
  \hline
  \tabincell{c}{Study}  & & \tabincell{c}{Sample \\ size}  & \tabincell{c}{Total\\ Cholesterol \\(mmol/L)}   & \tabincell{c}{LDL-C \\ (mmol/L)}  & \tabincell{c}{HDL-C \\(mmol/L)}   & \tabincell{c}{Triglycerides\\ (mmol/L)} \\   \hline   

   \multirow{2}{*}{\citet{bonnet2007}}    & \text{Case}   & 12 & 6.1(5.8-6.3)   & 4.1(3.7-4.6)   & 0.9(0.8-1.1)   & 2.0(1.1-3.3)   \\
                                          & \text{Control}& 9  & 6.4(6.1-7.7)   & 3.9(3.7-4.8)   & 1.0(0.8-1.1)   & 3.2(2.1-4.4)   \\
   \hline
   \multirow{2}{*}{\citet{calmy2010}}     & \text{Case}   & 10 & 5.6(4.5-6.4)   & 2.8(2.4-3.3)   & 1.0(0.9-1.3)   & 3.9(2.0-6.2)   \\
                                          & \text{Control}& 12 & 5.6(4.6-6.2)   & 3.5(2.5-4.1)   & 1.1(0.81-1.2)  & 2.3(1.5-3.5)   \\
   \hline
   \multirow{2}{*}{\citet{eckard2014}}    & \text{Case}   & 67 & \text{NS}      & 2.48(1.96-2.77)& \text{NS}      & \text{NS}      \\
                                          & \text{Control}& 69 & \text{NS}      & 2.50(1.99-3.13)& \text{NS}      & \text{NS}      \\
   \hline
   \multirow{2}{*}{\citet{funderburg2015}}& \text{Case}   & 72 & \text{NS}      & \text{NS}      & 1.21(0.98-1.49)& \text{NS}      \\
                                          & \text{Control}& 75 & \text{NS}      & \text{NS}      & 1.19(0.96-1.47)& \text{NS}      \\
   \hline
   \multirow{2}{*}{\citet{ganesan2011}}   & \text{Case}   & 22 & 4.34(3.72-4.45)& 2.50(2.25-2.82)& \text{NS}      & \text{NS}      \\
                                          & \text{Control}& 22 & 4.34(3.72-4.45)& 2.50(2.25-2.82)& \text{NS}      & \text{NS}      \\
   \hline
   \multirow{2}{*}{\citet{hurlimann2006}} & \text{Case}   & 29 & 6.4(6.0-7.4)   & 3.7(2.8-4.2)   & 1.2(1.1-1.6)   & 3.0(2.1-4.0)   \\
                                          & \text{Control}& 29 & 6.4(6.0-7.4)   & 3.7(2.8-4.2)   & 1.2(1.1-1.6)   & 3.0(2.1-4.0)   \\
   \hline
   \multirow{2}{*}{\citet{lo2015}}        & \text{Case}   & 17 & 5.14$\pm$0.98  & 3.20$\pm$0.95  & 1.34$\pm$0.50  & 1.36(1.10-2.31)\\
                                          & \text{Control}& 20 & 4.97$\pm$0.70  & 3.23$\pm$0.83  & 1.31$\pm$0.39  & 1.28(1.04-1.53)\\
   \hline
   \multirow{2}{*}{\citet{stein2004}}     & \text{Case}   & 20 & 5.58$\pm$0.40  & 3.47$\pm$0.32  & 0.94$\pm$0.07  & 3.78$\pm$0.67 \\
                                          & \text{Control}& 20 & 5.58$\pm$0.40  & 3.47$\pm$0.32  & 0.94$\pm$0.07  & 3.78$\pm$0.67 \\
   \hline
   \multirow{2}{*}{\citet{nakanjako2015}} & \text{Case}   & 15 & \text{NS}      & 3.1(2.2-4.9)   & 1.7(1.6-1.8)   & 1.6(1.1-2.4)   \\
                                          & \text{Control}& 15 & \text{NS}      & 4.9(2.4-6.7)   & 1.7(1.5-2.0)   & 2.0(1.4-3.2)   \\
   \hline
   \multirow{2}{*}{\citet{montoya2012}}   & \text{Case}   & 51 & \text{NS}      & 2.63(2.20-3.28)& \text{NS}      & \text{NS}      \\
                                          & \text{Control}& 53 & \text{NS}      & 2.53(2.30-3.10)& \text{NS}      & \text{NS}      \\
   \hline
   \multirow{2}{*}{\citet{moyle2001}}     & \text{Case}   & 14 & 7.5(6.7-8.3)   & 4.65(4.1-5.2)  & 0.94(0.79-1.08)& 3.96(2.84-6.52)\\
                                          & \text{Control}& 13 & 7.4(6.8-7.9)   & 4.68(3.89-5.47)& 0.87(0.72-1.02)& 4.06(2.20-5.97)\\
   \hline
   \multirow{2}{*}{\citet{mallon2006}}     & \text{Case}   & 14 & 7.6$\pm$1.7    & \text{NS}      & 1.1$\pm$0.4    & 3.8$\pm$4.1   \\
                                          & \text{Control}& 17 & 7.6$\pm$1.4    & \text{NS}      & 1.1$\pm$0.4    & 4.9$\pm$7.8   \\
  \hline
  \multicolumn{7}{l}{$\cdot$ Observations are expressed as \emph{mean $\pm$ SD} or \emph{median (interquartile range)}.}\\
  \multicolumn{7}{l}{$\cdot$ \emph{NS} indicates the information is not specified in the original study.}
\end{tabular}
\end{center}
\end{table}

\

\subsubsection{Results of symmetry test and meta-analysis}
\noindent
The test results are reported in Table \ref{table_6}.  For total cholesterol level, both case samples and control samples in study \citet{ganesan2011} have p-value less than 0.05, i.e. there is significant evidence to reject the symmetry null hypothesis for this study.  For LDL-C, the control samples in study \citet{montoya2012} has p-value less than 0.05.  For HDL-C, both case samples and control samples in study \citet{hurlimann2006} have small p-values ($p=0.024$) and for triglycerides, all the studies have non-significant test results.

\renewcommand\arraystretch{1.5}            
\begin{table}[htp]\footnotesize
\tabcolsep 4pt \caption{\small Results of normality test on reported data from \citet{banach2016}} \label{table_6} 
\begin{center}
\begin{tabular}{@{\extracolsep{\fill}}c|l|ccccc}  
  \hline
  \tabincell{c}{Study}  & & \tabincell{c}{Sample \\ size}  & \tabincell{c}{Total\\ Cholesterol}   & \tabincell{c}{LDL-C}  & \tabincell{c}{HDL-C}   & \tabincell{c}{Triglycerides} \\   \hline   

   \multirow{2}{*}{\citet{bonnet2007}}    & \text{Case}   & 12 & -0.450(0.653)   & 0.250(0.802)  & 0.750(0.453)   & 0.409(0.682)   \\
                                          & \text{Control}& 9  & 1.168(0.243)    & 1.190(0.234)  & -0.623(0.533)  & 0.081(0.935)   \\
   \hline
   \multirow{2}{*}{\citet{calmy2010}}     & \text{Case}   & 10 & -0.316(0.752)   & 0.223(0.824)  & 1.002(0.316)   & 0.191(0.849)   \\
                                          & \text{Control}& 12 & -0.563(0.574)   & -0.563(0.574) & -1.097(0.273)  & 0.450(0.653)   \\
   \hline
   \multirow{2}{*}{\citet{eckard2014}}    & \text{Case}   & 67 & NS             & -1.672(0.095)  & NS             & NS      \\
                                          & \text{Control}& 69 & NS             & 0.629(0.529)   & NS             & NS      \\
   \hline
   \multirow{2}{*}{\citet{funderburg2015}}& \text{Case}   & 72 & NS             & NS             & 0.599(0.549)   & NS      \\
                                          & \text{Control}& 75 & NS             & NS             & 0.612(0.540)   & NS      \\
   \hline
   \multirow{2}{*}{\citet{ganesan2011}}   & \text{Case}   & 22 & -2.254(0.024)  & 0.396(0.692)   & NS             & NS      \\
                                          & \text{Control}& 22 & -2.254(0.024)  & 0.396(0.692)   & NS             & NS      \\
   \hline
   \multirow{2}{*}{\citet{hurlimann2006}} & \text{Case}   & 29 & 1.613(0.107)   & -1.075(0.282)  & 2.258(0.024)   & 0.198(0.843)   \\
                                          & \text{Control}& 29 & 1.613(0.107)   & -1.075(0.282)  & 2.258(0.024)   & 0.198(0.843)   \\
   \hline
   \multirow{2}{*}{\citet{lo2015}}        & \text{Case}   & 17 & NS             & NS             & NS             & 1.585(0.113)\\
                                          & \text{Control}& 20 & NS             & NS             & NS             & 0.062(0.950)\\
   \hline
   \multirow{2}{*}{\citet{stein2004}}     & \text{Case}   & 20 & NS             & NS             & NS             & NS \\
                                          & \text{Control}& 20 & NS             & NS             & NS             & NS \\
   \hline
   \multirow{2}{*}{\citet{nakanjako2015}} & \text{Case}   & 15 & NS             & 0.860(0.390)   & 5.73$\times 10^{-15}$(1.000) & 0.596(0.551)   \\
                                          & \text{Control}& 15 & NS             & -0.420(0.674)  & 0.516(0.606)                 & 0.860(0.390)   \\
   \hline
   \multirow{2}{*}{\citet{montoya2012}}   & \text{Case}   & 51 & NS             & 1.039(0.299)   & NS             & NS      \\
                                          & \text{Control}& 53 & NS             & 2.213(0.027)   & NS             & NS      \\
   \hline
   \multirow{2}{*}{\citet{moyle2001}}     & \text{Case}   & 14 & 0.000(1.000)   & 0.000(1.000)   & -0.085(0.932)  & 0.969(0.333)\\
                                          & \text{Control}& 13 & -0.215(0.830)  & 2.366(0.018)   & 0.000(1.000)   & 0.031(0.975)\\
   \hline
   \multirow{2}{*}{\citet{mallon2006}}     & \text{Case}   & 14 & NS             & NS             & NS             & NS   \\
                                          & \text{Control}& 17 & NS             & NS             & NS             & NS   \\
  \hline
  \multicolumn{7}{l}{$\cdot$ Results are expressed as \emph{Test-statistic (p-value)}.}\\
  \multicolumn{7}{l}{$\cdot$ \emph{NS} indicates the original data do not need to conduct the symmetry test.}\\
  \multicolumn{7}{l}{$\cdot$ \emph{p-value}$<0.05$ indicates the underlying data is not symmetric.}
\end{tabular}
\end{center}
\end{table}

 \begin{figure}[htp]
 \begin{center}
 \subfigure[Total cholesterol]{
 \label{fig:case_2_TC_forest}
 \includegraphics[width=6.2in]{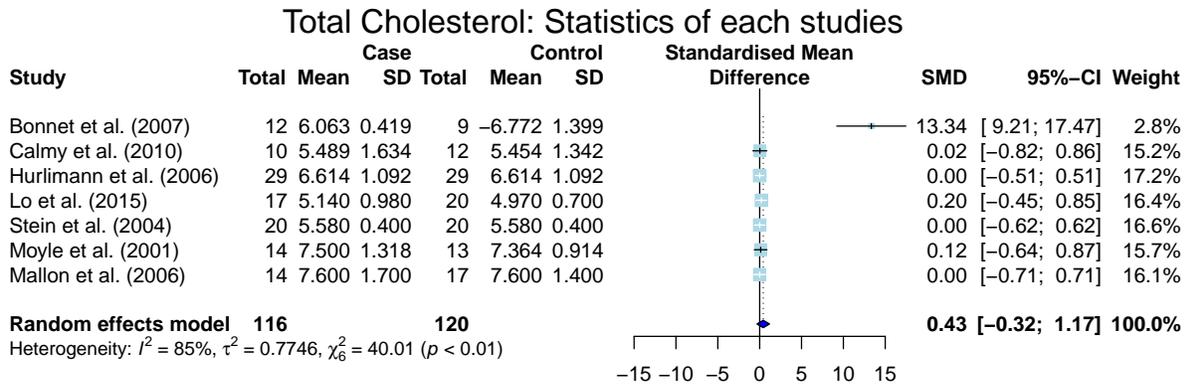}}
 \subfigure[LDL-C]{
 \label{fig:case_2_LDLC_forest}
 \includegraphics[width=6.2in]{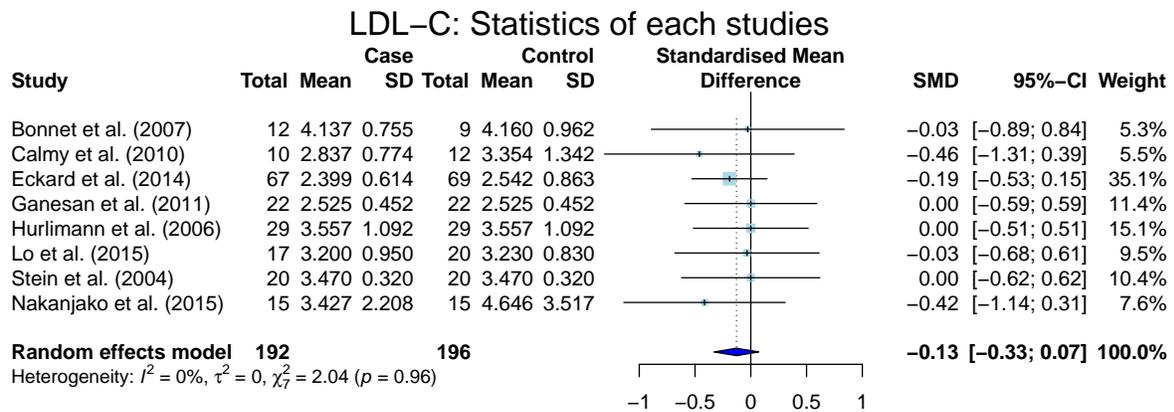}}
  \end{center}
 \caption{Forest plots of effects in plasma concentrations of total cholesterol (\ref{fig:case_2_TC_forest}), LDL-C (\ref{fig:case_2_LDLC_forest}) in HIV-infected patients.}
 \label{fig:case_2_forest_plots_1}
 \end{figure}

 \begin{figure}[htp]
 \begin{center}
  \subfigure[HDL-C]{
 \label{fig:case_2_DHLC_forest}
 \includegraphics[width=6.2in]{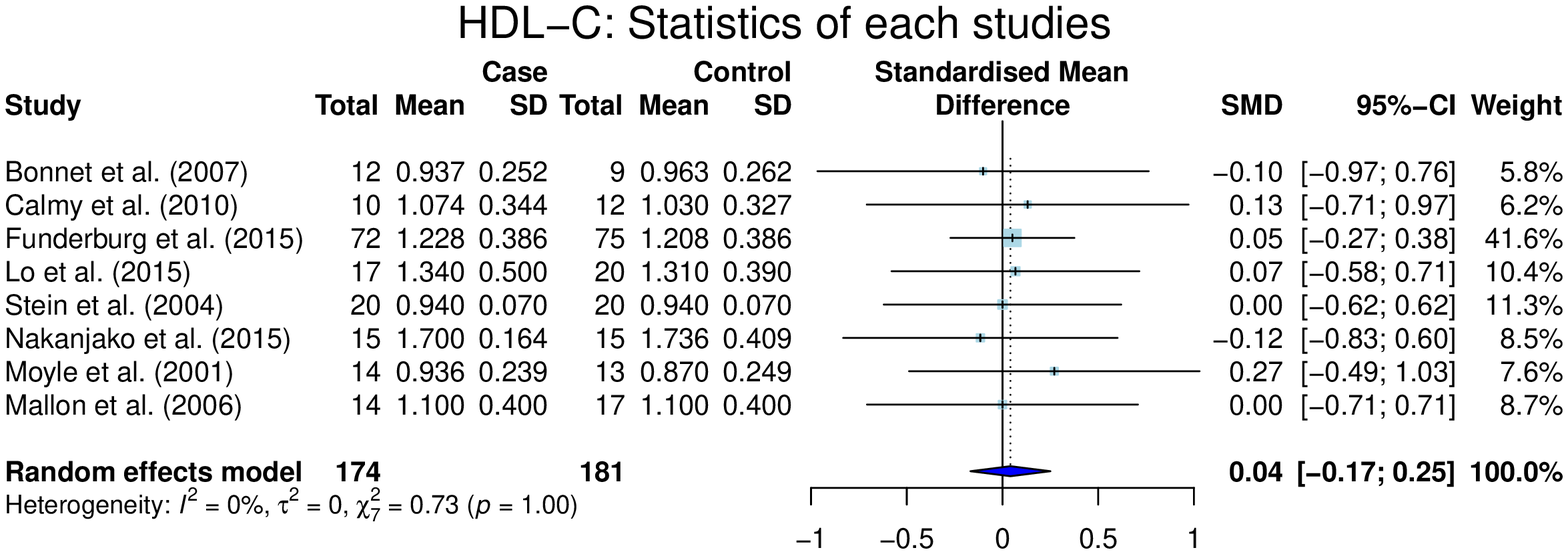}}
  \subfigure[Triglycerides]{
 \label{fig:case_2_trigly_forest}
 \includegraphics[width=6.2in]{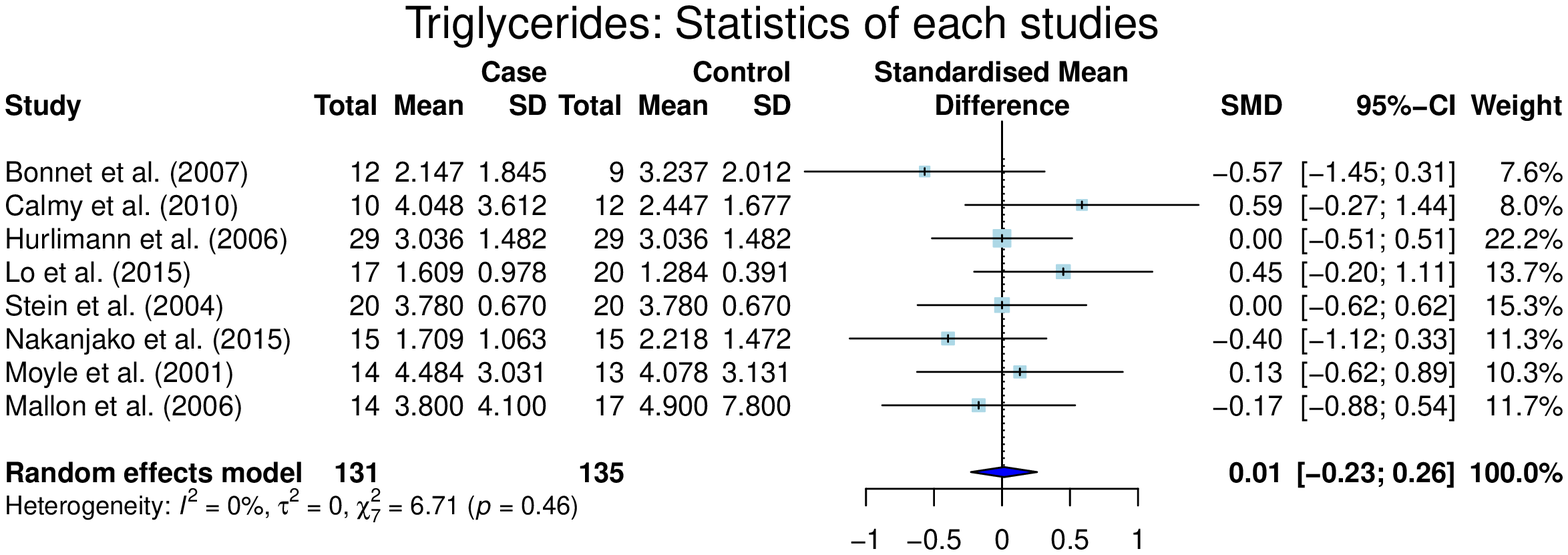}}
 \end{center}
 \caption{Forest plots of effects in plasma concentrations of HDL-C (\ref{fig:case_2_DHLC_forest}) and triglycerides (\ref{fig:case_2_trigly_forest}) in HIV-infected patients.}
 \label{fig:case_2_forest_plots_2}
 \end{figure}

Similar to Section 4.1.2, after removing the studies with significant test results (i.e. underlying data is more likely to be skewed), we obtain the following four forest plots of effects of statin therapy on total cholesterol, LDL-C, HDL-C and triglycerides levels in HIV-infected patients, respectively.  Likewise, we only provide overall results in this section due to demonstration purpose and will not further proceed sensitivity analysis and publication bias analysis.
The pooled results showed that there was very high heterogeneity among studies for total cholesterol level ($Q=40.01, I^{2}=85\%$ and $p<0.01$) and nearly no heterogeneity among studies for the other three plasma concentration levels ($I^{2}=0\%$ and $p>0.05$).
The overall effect sizes imply that by using statin therapy, HIV-infected patients may have moderate growth in total cholesterol level (SMD$=0.43$ with $95\%$ CI from -0.32 to 1.17), slight growth in HDL-C and triglycerides levels (SMD$=0.04$ with $95\%$ CI from -0.17 to 0.25 for HDL-C; SMD$=0.01$ with $95\%$ CI from -0.23 to 0.26 for triglycerides) and, slight reduction in LDL-C level (SMD$=-0.13$ with $95\%$ CI from -0.33 to 0.07).
Note that the pooled conclusions we obtained are different from \citet{banach2016}. In the original article, the authors concluded that there are significant reductions in both total cholesterol and LDL-C levels.  In contrast, we found that total cholesterol level has a moderate growth trend and LDL-C level only has a slight reduction.  As for HDL-C and triglycerides levels, nearly zero pooled effect sizes declared that these two plasma concentration levels do not have significant change in HIV-infected patients.  The opposite results we obtained indeed imply that it is essential to conduct a symmetry test on the selected studies before estimating the sample mean and standard deviation.  If the included studies are blindly used to estimate the sample mean and standard deviation without considering the symmetry of original data, it may increase computational error and eventually lead to an inaccurate conclusion.

\

\subsection{Case study of exploring the impact of statin therapy on plasma MMP-3, MMP-9 and TIMP-I concentrations}
\noindent
In this section, we will discuss the third real case as introduced in \citet{ferretti2017}.  This data is a meta-analysis about whether statin therapy has influence on plasma MMP-3, MMP-9 and TIMP-I concentration levels.  Total 10 studies were selected to conduct the systematic review.  Note that in \citet{ferretti2017}, the authors compared the net changes (mean difference) in measurements between pretreatment and posttreatment.  Since we tend to provide a demonstration for the usage of our proposed test statistics, we will not compute the net changes as in \citet{ferretti2017}.  Instead, the follow-up measurements, i.e. posttreatment measurements are used to conduct the test and further analysis.  Note also that in the original data set, study \citet{hanefeld2007} does not record any information of the control set and hence, we will remove it from the following analysis.

\

\subsubsection{Data description}
\noindent
The basic information of the selected data is reported in Table \ref{table_7}.  Similar to the previous two cases, for plasma MMP-9, MMP-3 and TIMP-I levels, data were recorded as the sample mean with standard deviation, or the sample median with interquartile range.  As a result, test statistic $T_{2}$ (Eq.(\ref{eq:final_test_s2})) will be used to test the symmetry of underlying data.  Likewise, if a study has p-value greater than 0.05, we regard it as appropriate to further conduct the meta-analysis.  If a study has p-value smaller than 0.05, it is recommended to be excluded from further analysis.

\renewcommand\arraystretch{1.5}            
\begin{table}[htp]\small
\tabcolsep 6pt \caption{\small Summary of included studies in \citet{ferretti2017}} \label{table_7} 
\begin{center}
\begin{tabular}{c|l|llll}  
  \hline
  \tabincell{c}{Study}  & & \tabincell{c}{Sample\\ size}  & \tabincell{c}{MMP-9\\(ng/mL)}   & \tabincell{c}{MMP-3\\(ng/mL)}  & \tabincell{c}{TIMP-I\\(ng/mL)}   \\   \hline

   \multirow{2}{*}{\citet{andrade2013}}         & \text{Case}   & 25 & 113$\pm$69           & NS            & 281$\pm$231   \\
                                                & \text{Control}& 8  & 147$\pm$89           & NS            & 354$\pm$287   \\
   \hline
   \multirow{2}{*}{\citet{koh2002}}             & \text{Case}   & 32 & 28(19-34)            & 16$\pm$14     & 74$\pm$23         \\
                                                & \text{Control}& 31 & 26(17-41)            & 18$\pm$17     & 86$\pm$26          \\
   \hline
   \multirow{2}{*}{\citet{mohebbi2014}}         & \text{Case}   & 21 & 164.95$\pm$126.68    & NS            & NS     \\
                                                & \text{Control}& 21 & 180.81$\pm$115.93    & NS            & NS     \\
   \hline
   \multirow{2}{*}{\citet{singh2008}$^{a}$}     & \text{Case}   & 23 & 10$\pm$6             & NS            & NS           \\
                                                & \text{Control}& 24 & 2$\pm$6              & NS            & NS            \\
   \hline
   \multirow{2}{*}{\citet{singh2008}$^{b}$}     & \text{Case}   & 22 & 9$\pm$6              & NS            & NS       \\
                                                & \text{Control}& 24 & 2$\pm$6              & NS            & NS       \\
   \hline
   \multirow{2}{*}{\citet{broch2014}}           & \text{Case}   & 36 & 243(106-367)         & NS            & NS      \\
                                                & \text{Control}& 35 & 354(162-467)         & NS            & NS      \\
   \hline
   \multirow{2}{*}{\citet{kalela2001}}          & \text{Case}   & 24 & 35.1$\pm$8.20        & NS            & NS   \\
                                                & \text{Control}& 26 & 40.4$\pm$25.30       & NS            & NS   \\
   \hline
   \multirow{2}{*}{\citet{leu2005}}             & \text{Case}   & 32 & 0.39$\pm$0.22        & NS            & NS     \\
                                                & \text{Control}& 19 & 0.42$\pm$0.22        & NS            & NS     \\
   \hline
   \multirow{2}{*}{\citet{nilsson2011}}         & \text{Case}   & 37 & 212(169-310)         & 21(16-28)     & 155(143-174)         \\
                                                & \text{Control}& 39 & 184(141-256)         & 20(14-24)     & 149(135-166)         \\
  \hline
  \multicolumn{6}{l}{$\cdot$ Observations are expressed as \emph{mean $\pm$ SD} or \emph{median (interquartile range)}.}\\
  \multicolumn{6}{l}{$\cdot$ \emph{NS} indicates the information is not specified in the original study.}
\end{tabular}
\end{center}
\end{table}

\

\subsubsection{Results of symmetry test and meta-analysis}
\begin{figure}[htp]
 \begin{center}
 \subfigure[MMP-9]{
 \label{fig:case_3_MMP9_forest}
 \includegraphics[width=6.2in]{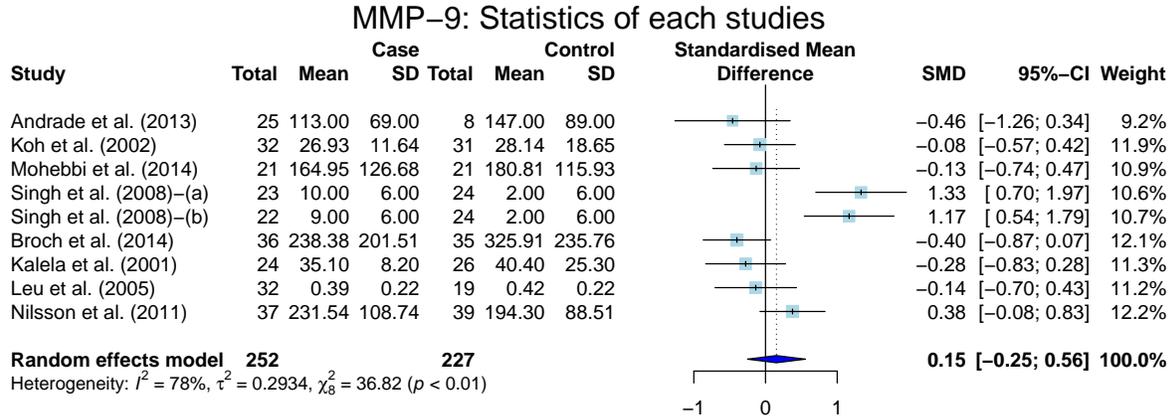}}
 \subfigure[MMP-3]{
 \label{fig:case_3_MMP3_forest}
 \includegraphics[width=6.2in]{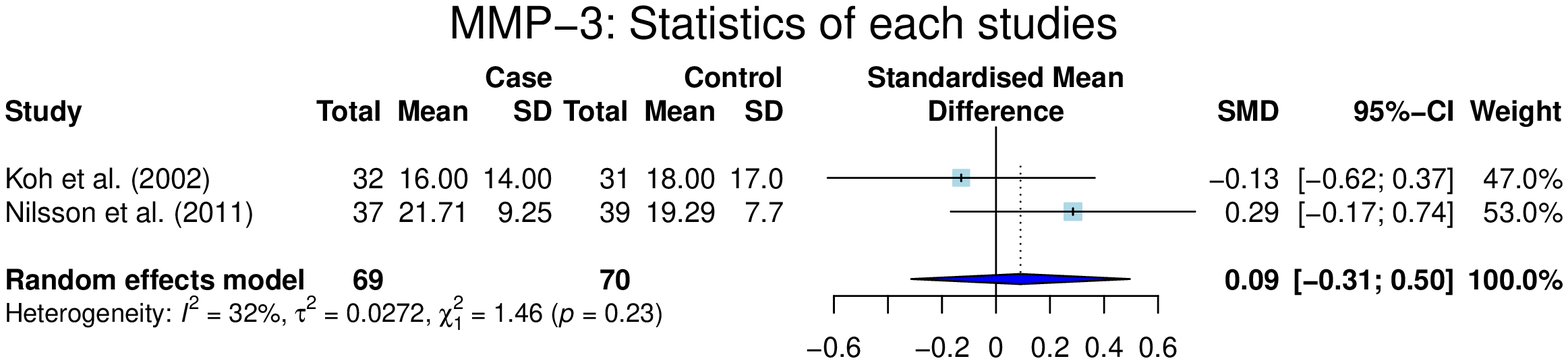}}
  \subfigure[TIMP-I]{
 \label{fig:case_3_TIMP_forest}
 \includegraphics[width=6.2in]{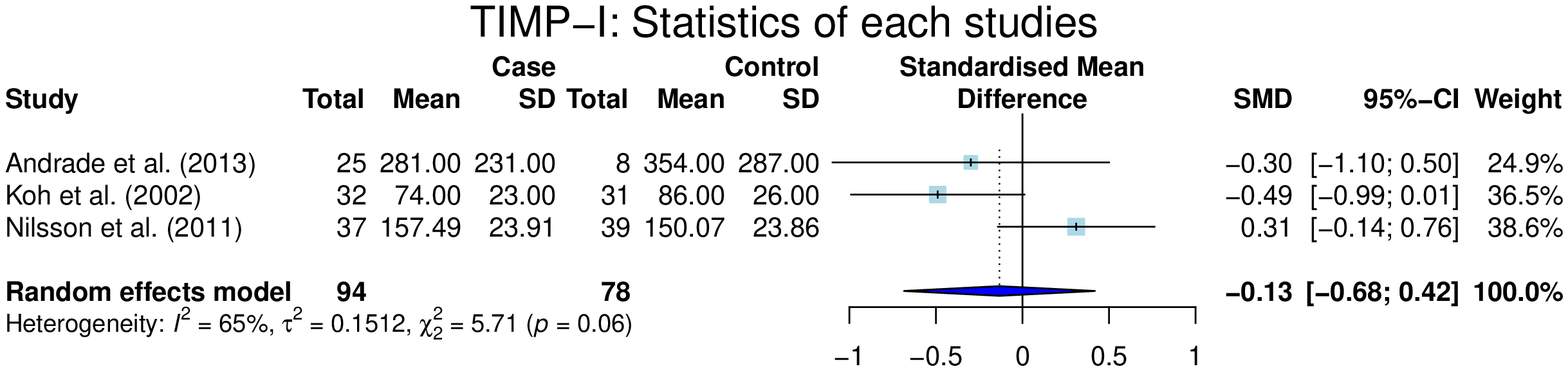}}
  \end{center}
 \caption{Forest plots of effects in follow-up plasma MMP-9 (\ref{fig:case_3_MMP9_forest}), MMP-3 (\ref{fig:case_3_MMP3_forest}) and TIMP-I (\ref{fig:case_3_TIMP_forest}) concentration levels in patients that use statin therapy.}
 \label{fig:case_3_forest_plots}
 \end{figure}

\noindent
The test results are provided in Table \ref{table_8}.  Since in the original data set, only 3 studies reported the sample median and interquartile range, Table \ref{table_8} only report the test results of these three studies.  Fortunately, the reported data in all three studies had passed the symmetry test (p-value greater than 0.05).  That is, we can apply the methods provided by \citet{luo2016} and \citet{Tong2014} to estimate the sample mean and standard deviation for these studies, respectively.  Similar to the previous two cases, we generated forest plots of overall effects from the statin therapy on patients' plasma MMP-3, MMP-9 and TIMP-I concentrations (Figures \ref{fig:case_3_MMP9_forest}, \ref{fig:case_3_MMP3_forest} and \ref{fig:case_3_TIMP_forest}).

\renewcommand\arraystretch{1.5}            
\begin{table}[htp]\small
\tabcolsep 7pt \caption{\small Results of symmetry test on reported data from \citet{ferretti2017}} \label{table_8} 
\begin{center}
\begin{tabular}{c|l|llll}  
  \hline
  \tabincell{c}{Study}  & & \tabincell{c}{Sample\\ size}  & \tabincell{c}{MMP-9\\(ng/mL)}   & \tabincell{c}{MMP-3\\(ng/mL)}  & \tabincell{c}{TIMP-I\\(ng/mL)}   \\   \hline

   \hline
   \multirow{2}{*}{\citet{koh2002}}             & \text{Case}   & 32 & -0.795(0.427)        & NS                 & NS         \\
                                                & \text{Control}& 31 & 0.976(0.329)         & NS                 & NS          \\
   \hline
   \multirow{2}{*}{\citet{broch2014}}           & \text{Case}   & 36 & -0.211(0.833)         & NS                & NS      \\
                                                & \text{Control}& 35 & -1.080(0.280)         & NS                & NS      \\
   \hline
   \multirow{2}{*}{\citet{nilsson2011}}         & \text{Case}   & 37 & 1.677(0.094)          & 0.716(0.474)      & 0.971(0.332)         \\
                                                & \text{Control}& 39 & 1.115(0.265)          & -0.884(0.376)     & 0.428(0.669)         \\
  \hline
  \multicolumn{6}{l}{$\cdot$ Results are expressed as \emph{Test-statistic (p-value)}.}\\
  \multicolumn{6}{l}{$\cdot$ \emph{NS} indicates the original data do not need to conduct the symmetry test.}\\
  \multicolumn{6}{l}{\footnotesize $\cdot$ \emph{p-value}$<0.05$ indicates the underlying data is not symmetric.}
\end{tabular}
\end{center}
\end{table}

Similar to the previous two case studies, we only compute overall effect sizes of the included data for demonstration purpose and will not conduct further analysis such as sensitivity analysis.  The effect sizes of each study and the overall results are shown in the three graphs of Figure \ref{fig:case_3_forest_plots}.  Based on these three forest plots, the pooled results imply that by using the statin therapy, patients' plasma MMP-9 and MMP-3 levels at the end of follow-up period may have a slightly increase (SMD$=0.15$ with $95\%$ CI from -0.25 to 0.56 for MMP-9 and SMD$=0.09$ with $95\%$ CI from -0.31 to 0.50 for MMP-3) than healthy individuals. On the other hand, patients' TIMP-I levels at the end of follow-up period may decrease very slightly (SMD$=-0.13$ with $95\%$ CI from -0.68 to 0.42).  Figure \ref{fig:case_3_MMP9_forest} also indicates that there might be high heterogeneity among studies for MMP-9 levels ($Q=35.41$, $I^{2}=77\%$ and $p<0.01$).  Note that in Figures \ref{fig:case_3_MMP3_forest} and \ref{fig:case_3_TIMP_forest}, the values of $I^{2}$ imply that there are low or moderate heterogeneity among studies for MMP-3 and TIMP-I levels, respectively ($I^{2}=32\%$ for MMP-3 and $I^{2}=65\%$ for TIMP-I).   However, the values of $Q$ statistics for these two plasma levels are very small and p-values are greater than 0.05 ($Q=1.46$ with $p=0.23$ for MMP-3; $Q=5.71$ with $p=0.06$ for TIMP-I), which imply that both MMP-3 and TIMP-I levels hardly have heterogeneity among selected studies.  Such conflict conclusion about heterogeneity between studies may be caused by the small number of studies included.  As a result, if researchers aim to obtain an accurate conclusion about the effects of statin therapy on patients' plasma MMP-3 and TIMP-I levels, more studies and information are needed.

Based on the above three real data analysis, it is obvious that our proposed test statistics are necessary for medical researchers, especially when we need to transform the intermediate summary statistics (median, minimum, maximum or interquartile range) to the sample mean and standard deviation.  We believe by conducting the proposed symmetry tests, researchers can efficiently find out and exclude the potential skewed data to reduce the errors in computing the effect sizes.

\

\section{Conclusion}
\noindent
Meta-analysis is a useful tool in evidenced-based medicine to statistically combine and analyze clinical results from two or more independent trials.
Researchers use different effect size measurements to statistically compare the effectiveness of some particular medicine or therapy.
Mean difference based effect sizes are used to analyze continuous data in clinical research.  To compute the mean difference measurements, the sample mean and standard deviation are two indispensable information.  However, most of the medical studies do not record these two statistics directly.  Instead, the summary measurements such as the sample median, minimum and maximum or interquartile range are more likely to be reported.  In this case, researchers have to transform the reported data into the sample mean and standard deviation.  Our major concern is that even for the optimal estimation methods of these two statistics (introduced by \citet{luo2016} and \citet{Tong2014}), the estimators are developed based on the symmetry assumption.  Thus we recommend that researchers should conduct a symmetry test for the underlying data before transform them into the sample mean and standard deviation, in order to improve the accuracy of the estimation.  As a result, we introduce test statistics based on the summary measurements for three most frequently appeared scenarios.

\renewcommand\arraystretch{1.5}
\begin{table} [htp]
\tabcolsep 8pt \caption{Summary table of the test statistics under three scenarios} \label{table_10} \vspace*{-8pt}
\begin{center}
\begin{tabular}{c|c|c}
  \hline
                                                & \text{Test statistic}             & \text{Coefficient}   \\   \hline
  \multirow{2}{*}{Scenario $\mathcal{S}_1$}     & \multirow{2}{*}{$T_{1}=\frac{\tau(n)\left(a+b-2m\right)}{b-a}$}
                                                & \multirow{2}{*}{\small{$\tau(n)=2\Phi^{-1}\left(\frac{n-0.375}{n+0.25}\right)\Big/\sqrt{\frac{\pi^2}{6\log(n)}+\frac{\pi}{n}}$}} \\
                                                & &  \\ \hline
  \multirow{2}{*}{Scenario $\mathcal{S}_2$}     & \multirow{2}{*}{$T_{2}=\frac{\varphi(n)\left(q_{1}+q_{3}-2m\right)}{q_{3}-q_{1}}$}
                                                & \multirow{2}{*}{\small{$\varphi(n)=1.09\sqrt{n}\Phi^{-1}\left(\frac{0.75n-0.125}{n+0.25}\right)$}}   \\
                                                & & \\  \hline
  \multirow{3}{*}{Scenario $\mathcal{S}_3$}     & \multirow{3}{*}{$T_{3}=\frac{\kappa(n)(a+b+q_{1}+q_{3}-4m)}{b-a+q_{3}-q_{1}}$}
                                                &\multirow{3}{*}{\large{$\kappa(n)=\frac{\left[2\Phi^{-1}\left(\frac{n-0.375}{n+0.25}\right)+2\Phi^{-1}\left(\frac{0.75n-0.125}{n+0.25}\right)\right]}{\sqrt{\frac{\pi^2}{6\log(n)} + \frac{10.5}{n}}}$}} \\
                                                & & \\
                                                & & \\
  \hline
\end{tabular}
\end{center}
\end{table}

For the three most popular scenarios, the corresponding test statistics and their coefficient functions are summarized in Table \ref{table_10}.  All these three test statistics are theoretically proved and have simple formulation which are easy for researchers to adopt in practice.  The simulation studies in Section 3 show that our proposed test statistics have statistical power close to one and for both scenarios, which indicates that the test statistics should have very good performance in detecting potential skewed data.  Moreover, to help researchers to make more convincing conclusions in medical research, we suggest a proper path to conduct the meta-analysis in Figure \ref{fig:meta_procedure} and the first step is to conduct the symmetry test via our test statistics first.
To further illustrate the usage of the newly proposed test statistics, we applied them on three different real meta-analysis in Section 4.  By conducting the symmetry test, the p-values show that some of the selected studies of these three cases should be excluded before carry on to meta-analysis.  As a result, the results of meta-analysis with skewed data removed may be different from the original conclusion.  In particularly, for the case study about the effects of statin therapy on plasma lipid concentrations in HIV-infected patients \citep{banach2016}, compare to the original article, we have opposite conclusion about the pooled effect sizes of total cholesterol level and LDL-C level.  Since the simulation studies indicate the reliability of the proposed test statistics, we expect that our test statistics and the recommended procedure in Figure \ref{fig:meta_procedure} can help researchers to obtain faithful conclusion in evidence-based medicine.

\newpage
\vskip 24pt
\bibliographystyle{apalike2}
\bibliography{overall_ref7}

\newpage
\appendix

\centerline{\large {\bf Appendix A: Some preliminary results}}

\vskip 12pt
\noindent
To derive the symmetry test for the three scenarios, we first present some preliminary results for the normal distribution and for the associated order statistics.
The normal distribution $N(\mu,\sigma^{2})$ is commonly used in statistics for data analysis.
Its probability density function (pdf) is given as
\begin{equation*}
\phi(x|\mu,\sigma^2)=\frac{1}{\sqrt{2\pi\sigma^{2}}}{\rm{exp}}\left\{-\frac{(x-\mu)^{2}}{2\sigma^{2}}\right\},
\end{equation*}
where $\mu$ is the mean value and $\sigma^{2}$ is the variance, or equivalently, $\sigma$ is the standard deviation.
For the normal distribution, $\mu$ is also known as the median and the mode.
When $\mu=0$ and $\sigma^{2}=1$, the distribution reduces to the standard normal distribution $N(0,1)$.
Let also $\Phi(\cdot)$ be the cumulative density function (cdf) of the standard normal distribution.
By symmetry, we have $\phi(z) = \phi(-z)$ and $\Phi(z) = 1-\Phi(-z)$.

To investigate the properties of the 5-number summary for the data, we introduce some theoretical results for the order statistics $Z_{(1)}\leq \cdots \leq Z_{(n)}$
of the random sample $\{Z_{1}, \ldots,Z_{n}\}$ from the standard normal distribution.
By symmetry, $Z_{(i)}$ and $-Z_{(n-i+1)}$ follow the same distribution,
and $(Z_{(i)},Z_{(j)})$ and $(Z_{(n-i+1)},Z_{(n-j+1)})$ follow the same joint distribution.
According to Arnold and Balakrishnan \citep{relations-order-statistics}, Chen \citep{Chenhung} and Ahsanullah et al. \citep{intro-to-order-stat}, we have the following two lemmas.

\vskip 12pt
\begin{lemma} \label{theorem:order_statistics}
Let $Z_{1}, \ldots,Z_{n}$ be a random sample of $N(0,1)$, and $Z_{(1)}\leq \cdots \leq Z_{(n)}$ be the ordered statistics $Z_{1}, \ldots, Z_{n}$.
Then
\begin{align*}
E(Z_{(i)})        &= -E(Z_{(n-i+1)}), \quad 1\leq i\leq n,\\
E(Z_{(i)}Z_{(j)}) &= E(Z_{(n-i+1)}Z_{(n-j+1)}), \quad 1\leq i\leq j\leq n.
\end{align*}
\end{lemma}

\vskip 12pt
\begin{lemma} \label{theorem:sample_quantile_property}
Let $Z_{1}, \ldots,Z_{n}$ be a random sample of $N(0,1)$, and $Z_{([np])}$ be the $p$th quantile of the sample, where $[np]$ denotes the integer part of $np$.
Let also $\Phi^{-1}(\cdot)$ be the inverse function of $\Phi(\cdot)$.
\begin{enumerate}[(1)]
\item[{\rm (i)}] For any $0<p<1$, we have
\begin{equation*}
\sqrt{n}(Z_{([np])}-\Phi^{-1}(p)) \xrightarrow{d} N\left(0,\frac{p(1-p)}{[\phi(\Phi^{-1}(p))]^{2}}\right), ~~~{\rm as}~ n\to\infty,
\end{equation*}
where $\xrightarrow{d}$ denotes convergence in distribution.
\item[{\rm (ii)}] For any $0<p_{1} <p_{2}<1$, as $n\to \infty$, $(Z_{([np_{1}])},Z_{([np_{2}])})$ follows asymptotically a bivariate normal distribution
with mean vector $(\Phi^{-1}(p_{1}),\Phi^{-1}(p_{2}))$ and covariance matrix $\Sigma =(\sigma_{ij})_{2\times 2}$, where $\sigma_{12}=\sigma_{21}$ and
           \begin{equation*}
           \sigma_{ij}=\frac{p_{i}(1-p_{j})}{n \phi(\Phi^{-1}(p_{i})) \phi(\Phi^{-1}(p_{j}))}, ~~~~~1 \leq i\leq j\leq 2.
           \end{equation*}
\end{enumerate}
\end{lemma}

\vskip 12pt
\begin{lemma} \label{theorem:extremum_property}
Let $Z_{1}, \ldots,Z_{n}$ be a random sample of $N(0,1)$ and $Z_{(1)}\leq Z_{(2)}\leq \cdots \leq Z_{(n)}$ be the corresponding order statistics.  According to Theorem 14 in \citet{ferguson1996},
the sample maximum $Z_{(n)}$ has the following limiting distribution:
\begin{equation*}
\sqrt{2\log (n)}Z_{(n)} -2\log(n) +\frac{1}{2}\log(4\pi\log(n)) \rightarrow Y,
\end{equation*}
where $Y\in G_{3}=\exp\left\{-e^{-z} \right\}$.  By symmery, the sample minimum also has the following limiting distribution:
\begin{equation*}
\sqrt{2\log (n)}Z_{(2)} +2\log(n) -\frac{1}{2}\log(4\pi\log(n)) \rightarrow -W,
\end{equation*}
where $W\in G_{3}$.  By Theorem 15 in \citet{ferguson1996}, the above two expressions converge jointly as $Y$ and $W$ are independent.  Hence, for the sample mid-range, $\frac{Z_{(1)}+Z_{(n)}}{2}$, satisfies,
\begin{equation*}
\sqrt{2\log(n)}\left(\frac{Z_{(1)}+Z_{(n)}}{2}\right) \rightarrow \frac{Y-W}{2}.
\end{equation*}
where $\frac{Y-W}{2}$ follows the logistic distribution $L(0,\frac{1}{2})$.

\end{lemma}

\vskip 30pt
\centerline{\large {\bf Appendix B: Theoretical results of proposed test statistics}}
\vskip 12pt
\noindent
Recall that in Section 3.2, $X_{1}, \ldots, X_{n}$ are defined as a random sample of size $n$ from the normal distribution $N(\mu,\sigma^{2})$,
and $X_{(1)}\leq \cdots \leq X_{(n)}$ are the ordered statistics of the sample.
Meanwhile, they can represented as $X_{i}=\mu+\sigma Z_{i}$ and $X_{(i)}=\mu+\sigma Z_{(i)}$ for $i=1, \ldots, n$.
By the three lemmas in Appendix A, we have the following theoretical results for the proposed estimators under the three scenarios, respectively.

\vskip 24pt
\begin{theorem} \label{lemma:test_S1}
	Under the null hypothesis, the test statistic $T$ for scenario $\mathcal{S}_{1}$ can be further written as
	\begin{equation*}
	T_0=\frac{a+b-2m}{\sigma\sqrt{\frac{\pi^2}{6\log(n)}+\frac{\pi}{n}}},
	\end{equation*}
	where $\sigma$ can be estimated by Eq. (\ref{eq:estimate_sigma_s1}).
\end{theorem}

\vskip 24pt
\noindent
{\bf Proof}. Recall the test statistic $T$ in (\ref{eq:test_statistic_s1}), i.e.
\begin{equation*}
T=\frac{a+b-2m}{{\rm SE}(a+b-2m)}.
\end{equation*}

By Lemma \ref{theorem:extremum_property}, we know that the sample mid-range for standard normal distributed data, $\left(Z_{(1)}+Z_{(n)}\right)/2$, has variance ${\rm Var}\left(Z_{(1)}+Z_{(n)}\right)=\pi^2/\left[6\log(n)\right]$ when $n$ is large.  By Lemma \ref{theorem:sample_quantile_property}, we can have the variance of the median for standard normal distributed data is $Var(Z_{([0.5n])})=\frac{\pi}{2n}$, when $n$ is large.  Hence, under the assumption in Section 3.2, for $a=\mu+\sigma Z_{(1)}$, $b=\mu+\sigma Z_{(n)}$ and $m=\mu+\sigma Z_{[0.5n]}$, we have:
\begin{equation*}
\begin{split}
{\rm Var}\left(a+b-2m\right)&={\rm Var}(a+b)+4{\rm Var}(m)-4{\rm Cov}(a+b,m)\\
                            &\approx \frac{\pi^2\sigma^2}{6\log(n)}+\frac{2\pi\sigma^2}{n}-4{\rm Cov}(a+b,m).
\end{split}
\end{equation*}
According to the simulation results of ${\rm Cov}(a+b,m)$, we found it very close to the variance of the sample median, i.e. ${\rm Var}(m)$.  Thus, we computed the ratio ${\rm Cov}(a+b,m)/{\rm Var}(m)$ and tended to use ${\rm Var}(m)$ to approximate the values of ${\rm Cov}(a+b,m)$.  Based on the simulation results, we figured out the ratio between these two terms is approximately equal to 0.5.  By plugging the this ratio between ${\rm Cov}(a+b,m)$ and ${\rm Var}(m)$ into the previous formula of the variance, we can obtain the following equation:
\begin{equation*}
\begin{split}
{\rm Var}\left(a+b-2m\right)&\approx \frac{\pi^2\sigma^2}{6\log(n)}+\frac{2\pi\sigma^2}{n} - 4\left(\frac{1}{2}\right)\left(\frac{\pi\sigma^2}{2n}\right)\\
                            &= \frac{\pi^2\sigma^2}{6\log(n)}+\frac{\pi\sigma^2}{n}.
\end{split}
\end{equation*}

Recall the test statistic $T$ in (\ref{eq:test_statistic_s1}), the denominator
\begin{scriptsize}{${\rm SE}\left(a+b-2m\right)=\sqrt{{\rm Var}\left(a+b-2m\right)}$}\end{scriptsize},
and from the above derivation, by plugging in the above results into (\ref{eq:test_statistic_s1}), we have,
\begin{equation*}
T_0=\frac{a+b-2m}{\sigma\sqrt{\frac{\pi^2}{6\log(n)}+\frac{\pi}{n}}},
\end{equation*}
is the test statistic under the null hypothesis $H_0$.
Substituting $\sigma$ in above equation by (\ref{eq:estimate_sigma_s1}), we can easily obtain the finalized test statistic (\ref{eq:final_test_s1})

\vskip 24pt
\begin{theorem} \label{lemma:test_S2}
	Under the null hypothesis, the test statistic $T$ for scenario $\mathcal{S}_{2}$ can be further written as
	\begin{equation*}
	T_0 = \frac{\sqrt{n}\left(q_{1}+q_{3}-2m\right)}{1.83\sigma},
	\end{equation*}
	where $\sigma$ can be estimated by Eq. (\ref{eq:estimate_sigma_s2}).
\end{theorem}

\vskip 24pt
\noindent
{\bf Proof}. Recall the test statistic $T$ in Eq. (\ref{eq:test_statistic_s2}),
\begin{equation*}
T=\frac{q_{1}+q_{3}-2m}{{\rm SE}(q_{1}+q_{3}-2m)}
\end{equation*}

According to the notations in Section 3.2, the sample first and third quartiles and the median are represented by $q_{1}=\mu+\sigma Z_{[0.25n]}$, $m=\mu+\sigma Z_{[0.5n]}$ and $q_{3}=\mu+\sigma Z_{[0.75n]}$.
By Lemma \ref{theorem:order_statistics} and \ref{theorem:sample_quantile_property}, when sample size $n$ is large, we have
      ${\rm Var}(Z_{([0.25n])})=\frac{0.25(0.75)}{n[\phi(\Phi^{-1}(0.25))]^{2}}\approx \frac{1.8568}{n}$,
      ${\rm Var}(Z_{([0.5n])})=\frac{\pi}{2n}$,
      ${\rm Var}(Z_{([0.75n])})\approx \frac{1.8568}{n}$,
      ${\rm Cov}(Z_{([0.25n])},Z_{[0.5n]})\approx \frac{0.9860}{n}$,
      ${\rm Cov}(Z_{([0.25n])},Z_{([0.75n])})\approx \frac{0.6189}{n}$.
Hence, we could easily obtain that
      \begin{equation*}
      \begin{split}
      {\rm Var}(q_{1})&={\rm Var}(q_{3})\approx1.8568\sigma^{2}/n,\qquad {\rm Var}(m)=\pi\sigma^{2}/2n,\\
      {\rm Cov}(q_{1},m)&\approx0.9860\sigma^{2}/n, \qquad {\rm Cov}(q_{1},q_{3})\approx0.6189\sigma^{2}/n.
      \end{split}
      \end{equation*}

As a result, under the null hypothesis $H_{0}$, the test statistic $T$ in (\ref{eq:test_statistic_s2}) can be simplified as
\begin{equation*}
\begin{split}
T_{0}&=\frac{q_{1}+q_{3}-2m}{\sqrt{{\rm Var}(q_{1}+q_{3}) - 4{\rm Cov}(q_{1}+q_{3},m) + 4{\rm Var}(m)}}\\
 &=\frac{q_{1}+q_{3}-2m}{\sqrt{[2{\rm Var}(q_{1}) + 2{\rm Cov}(q_{1},q_{3})] - 4[2{\rm Cov}(q_{1},m)] + 4{\rm Var}(m)}}\\
 &=\frac{q_{1}+q_{3}-2m}{\sqrt{\frac{\sigma^2}{n}\left[2(1.8568) + 2(0.6189) - 8(0.9860) + 2\pi \right]}}\\
 &\approx \frac{\sqrt{n}\left(q_{1}+q_{3}-2m\right)}{1.83\sigma}.
\end{split}
\end{equation*}
Use (\ref{eq:estimate_sigma_s2}) to estimate the unknown parameter $\sigma$, we can obtain the finalized test statistic as (\ref{eq:final_test_s2}).

\vskip 24pt
\begin{theorem} \label{lemma:test_S3}
	Under the null hypothesis, the test statistic $T$ scenario $\mathcal{S}_{3}$ can be further written as
	\begin{equation*}
	T_{0}= \frac{a+b+q_{1}+q_{3}-4m}{\sigma\sqrt{\frac{\pi^2}{6\log(n)} + \frac{10.14}{n}}},
	\end{equation*}
	where $\sigma$ can be estimated by Eq. (\ref{eq:estimate_sigma_s3}).
\end{theorem}

\vskip 24pt
\noindent
{\bf Proof}. Recall the test statistic $T$ in (\ref{eq:test_statistic_s3}), i.e.
\begin{equation*}
T=\frac{a+b+q_{1}+q_{3}-4m}{{\rm SE}(a+b+q_{1}+q_{3}-4m)}.
\end{equation*}

By Lemma 1, we have for standard normal distribution, \begin{footnotesize}{${\rm Var}(Z_{([0.25n])})={\rm Var}(Z_{([0.75n])})$}\end{footnotesize},
\begin{footnotesize}{${\rm Cov}(Z_{([0.25n])},Z_{([0.5n])})={\rm Cov}(Z_{([0.75n])},Z_{([0.5n])})$}\end{footnotesize}
and \begin{footnotesize}{${\rm Cov}(Z_{(1)}+Z_{(n)},Z_{([0.25n])})={\rm Cov}(Z_{(1)}+Z_{(n)},Z_{([0.75n])})$}\end{footnotesize}.
Also by Theorem \ref{lemma:test_S1}, we can easily obtain that ${\rm Cov}(a+b,m)\approx 0.5{\rm Var}(m)$ when $n$ is large.  Similar as Theorem \ref{lemma:test_S1}, by simulation, we found that the covariance term ${\rm Cov}(a+b,q_1)$ is very close to the variance of first quartile, i.e. ${\rm Var}(q_1)$.  As a result, we compute the ratio between ${\rm Cov}(a+b,q_1)$ and ${\rm Var}(q_1)$ to seek an approximation of ${\rm Cov}(a+b,q_1)$, with respect to ${\rm Var}(q_1)$.  Eventually, we figured out that the ratio is approach to 0.45 as $n$ increases.
Consequently, by plugging in the approximation equation of ${\rm Cov}(a+b,q_1)$ and, using Lemma \ref{theorem:sample_quantile_property} and \ref{theorem:extremum_property},we have
\begin{small}
\begin{equation*}
\begin{split}
{\rm Var}(a+b+q_{1}+q_{3}-4m)&={\rm Var}(a+b)+2{\rm Var}(q_{1})+16{\rm Var}(m) \\
                             & \qquad\quad +4{\rm Cov}(a+b,q_{1})-8{\rm Cov}(a+b,m)-16{\rm Cov}(q_{1},m)   \\
                             &\approx {\rm Var}(a+b)+2{\rm Var}(q_{1})+16{\rm Var}(m) \\
                             & \qquad\quad +4(0.45){\rm Var}(q_{1})-8(0.5){\rm Var}(m)-16{\rm Cov}(q_{1},m)   \\
                             &\approx \frac{\pi^2\sigma^2}{6\log(n)} + \frac{3.72\sigma^{2}}{n} + \frac{8\pi\sigma^2}{n} + \frac{3.348\sigma^{2}}{n} - \frac{2\pi\sigma^2}{n} - \frac{15.78\sigma^2}{n} \\
                             &\approx \frac{\pi^2\sigma^2}{6\log(n)} + \frac{7.068\sigma^{2}}{n} + \frac{6\pi\sigma^2}{n} - \frac{15.78\sigma^{2}}{n} \\
                             &\approx \frac{\pi^2\sigma^2}{6\log(n)} + \frac{10.14\sigma^{2}}{n}
\end{split}
\end{equation*}
\end{small}

Eventually, under the null hypothesis $H_{0}$, the test statistic in (\ref{eq:test_statistic_s3}) can be simplified as
\begin{equation*}
T_{0}= \frac{a+b+q_{1}+q_{3}-4m}{\sigma\sqrt{\frac{\pi^2}{6\log(n)} + \frac{10.14}{n}}}.
\end{equation*}
After substitute the unknown parameter $\sigma$ by the estimated $\hat{\sigma}$ in (\ref{eq:estimate_sigma_s3}), we can obtain the finalized test statistic $T_{3}$ as in (\ref{eq:final_test_s3}).

\end{document}